\documentclass[a4paper,showpacs,twocolumn,amsmath,amssymb,floatfix,prb,preprintnumbers,footinbib,eqsecnum, unsortedaddress]{revtex4}

\usepackage[pdftex]{graphicx}
\usepackage{dcolumn}% Align table columns on decimal point
\usepackage{bm}% bold math
\usepackage{dsfont}
\usepackage{color}

\usepackage[english]{babel}

\newcommand{\ee}{\mathrm{e}}
\newcommand{\ii}{\mathrm{i}}

\newcommand{\qq}{\mathcal{Q}}
\newcommand{\vf}{v_{\text{F}}}
\newcommand{\s}{\text{S}}
\newcommand{\D}{\text{D}}
\newcommand{\T}{\text{T}}

\newcommand{\no}[1]{:\!#1\!:}

\usepackage{color}
\usepackage[colorlinks=true ,citecolor=blue]{hyperref}

\newcommand{\kp}{\emph{\textbf{k}}$\cdot$\emph{\textbf{p}}\ }

\newcommand{\comment}[1]{}

\begin{document}

\title{Defect-induced multicomponent electron scattering in single-walled carbon nanotubes}
\author{Dario Bercioux}
\email{dario.bercioux@frias.uni-freiburg.de}
\affiliation{Freiburg Institute for Advanced Studies, Albert-Ludwigs-Universit\"at, D-79104 Freiburg, Germany}
\affiliation{Physikalisches Institut, Albert-Ludwigs-Universit\"at, D-79104 Freiburg, Germany}
\author{Gilles Buchs}
\email{g.buchs@tudelft.nl}
\affiliation{Kavli Institute of Nanoscience, TU-Delft, P.O. Box 5046, 2600 GA Delft, The Netherlands}
\author{Hermann Grabert}
\affiliation{Freiburg Institute for Advanced Studies, Albert-Ludwigs-Universit\"at, D-79104 Freiburg, Germany}
\affiliation{Physikalisches Institut, Albert-Ludwigs-Universit\"at, D-79104 Freiburg, Germany}
\author{Oliver Gr\"oning}
\affiliation{EMPA Swiss Federal Laboratories for Materials Testing and Research, nanotech@surfaces, Feuerwerkerstr. 39, CH-3602 Thun, Switzerland}

\begin{abstract}
We present a detailed comparison between theoretical predictions on electron scattering processes in metallic single-walled carbon nanotubes with defects and experimental data obtained by scanning tunneling spectroscopy of Ar$^+$ irradiated nanotubes. To this purpose we first develop  a formalism for studying quantum transport properties of defected nanotubes in   presence of source and drain contacts and an STM tip. The formalism is based on a field theoretical approach describing low-energy electrons. We account for the lack of translational invariance induced by defects within the so called \emph{extended} \kp approximation. 
The theoretical model reproduces the features of the particle-in-a-box-like states observed experimentally.   Further, the comparison between theoretical and experimental   Fourier-transformed local density of state maps yields clear signatures for inter- and intra-valley electron scattering processes depending on the tube chirality.
\end{abstract}

\date{\today}


% 73.63.Fg	Nanotubes
% 73.22.-f	Electronic structure of nanoscale materials: clusters, nanoparticles, nanotubes, and nanocrystals
% 68.37.Ef Scanning tunneling microscopy (including chemistry induced with STM)

\date{\today}

\maketitle

\section{Introduction}\label{sec:one}
In the last two decades, carbon nanotubes  discovered in 1991~\cite{Iijima:1991} and in particular Single-Walled Carbon Nanotube (SWNT) that emerged shortly~\cite{IIJIMA:1993.737,BETHUNE:1993.605} after have attracted a continuously increasing interest from the solid states physics community due to their outstanding electronic and mechanical properties.~\cite{saito:1998.icp, tomanek:1999.ka, ando:74.777, charlier:677, Steele:2009p1103, Lassagne:2009p1107} A single-walled carbon nanotube (SWNT) can be regarded as a strip of graphene rolled into a hollow cylinder with diameters typically in the order of one nanometer and lengths ranging from a few microns to a few centimeters.~\cite{charlier:677} Because of the very small diameter, a SWNT is generally considered as a quasi-one-dimensional system. SWNTs can be either metallic or semiconducting with energy bandgaps varying from zero to above 1~eV, depending on their geometrical structure.~\cite{charlier:677}

The low energy spectrum of metallic SWNT comprises four branches with a linear dispersion close to the charge neutrality point (CNP). Depending on the chirality, the corresponding electronic states can be degenerate in the orbital momentum. These states are symmetric with respect to the energy plane through the CNP implying electron-hole symmetry. \cite{JarilloHerrero:2004p429} Other dispersive, energy symmetric electronic states are opening at higher energy, with a gap~$\ge1$eV. All these electronic states can be obtained theoretically from a \kp method.~\cite{ando:74.777} This approach is particularly useful for studying transport properties.

An ideal metallic SWNT | defectless, charge-neutral and infinitely long | can be regarded as a quantum wire with four propagating one-dimensional conduction channels, \emph{i.e.} two left- and two right-movers.~\cite{note:one} This leads to a quantized conductance at zero temperature of $4e^2/h$, since each left- or right-moving channel is two-fold degenerate in spin.

In practice, devices with pure and clean SWNTs are hardly achievable. Their structure is modified by random potentials arising from adsorbates, defects, inhomogeneous substrate interactions, Schottky barriers at contacts etc.. These perturbations lead to a finite backscattering probability and hence an experimentally measured conductance below the theoretical maximum of $4e^2/h$. The scattering mean-free path $\ell_\text{p}$ of SWNTs has been experimentally determined and is of the order of $\ell_\text{p}\sim 1~\mu$m. It is usually longer in metallic SWNTs and shorter in semiconducting ones.~\cite{charlier:677}

SWNTs constitute an ideal platform to study many body interactions in one-dimensional conductors. A field theoretical approach based on the Tomonaga-Luttinger liquid model~\cite{TOMONAGA:1950.544,LUTTINGER:1963.1154} has been proposed a decade ago~\cite{Egger:1997.5082,Egger:1998.281,PhysRevLett.79.5086,PhysRevLett.83.5547} in order to describe interaction effects in isolated single-walled and multi-walled carbon nanotubes. Some experiments have reported on power-law behavior of the conductance as function of the applied voltage and/or the temperature for SWNTs~\cite{Bockrath:1999.598,Yao:1999.273} and for metallic junctions of crossed SWNTs~\cite{PhysRevLett.92.216804} with exponents in the range of theoretical predictions.

Transport experiments on SWNTs cover several distinct regimes characterized by the number of external electrodes and the quality of the contacts between the electrodes and the tube.
Source and drain electrodes drive an electric current through a contacted SWNT. These electrodes can be either weakly or strongly coupled to the nanotube. In the first case | the so-called Coulomb blockade regime | effects of Coulomb charging play an important role.~\cite{Postma:293-76} Oppositely, for strong coupling | the so-called Fabry-P\'erot regime~\cite{pugnetti:035121} | the transmissivity of the contacts is close to unity~\cite{Liang:2001p2177,wu:156803,herrmann:156804} therefore minimizing the effects of Coulomb charging. The use of a back gate permits to shift the position of the SWNT bands relative to the Fermi level in order to probe the conduction channels as a function of energy. In a typical strongly coupled two-terminal device, the SWNT differential conductance | the derivative of the average current with respect to the applied voltage | shows an oscillatory behavior  as a function of the  source-drain and gate voltages. These oscillations  can be easily interpreted in terms of a Fabry-P\'erot electron resonator model:~\cite{Liang:2001p2177}  at low energy the electrons propagate ballistically between the two contacts and the oscillation pattern of the differential conductance is the result of coherent interference.

Detailed investigations of transport and structural properties of SWNTs can be achieved by means of scanning probe techniques as, \emph{e.g.}, scanning tunneling microscopy/spectroscopy (STM/STS) or scanning gate microscopy (SGM). In both cases, an atomically sharp metallic tip is brought in close proximity to the nanotube. In this configuration, by applying a bias voltage it is possible to generate a tunneling current  between the conductive sample and the tip.
This current is kept constant by means of a feedback loop whose output signal adjusts the vertical $z$-position of the tip as a function of the $x$-$y$ position. The recorded $z(x,y)$ signal reflects a constant current contour map | it can be interpreted, in a first approximation, as the topography of the sample. In STS, the tip is stopped at a given position and the feedback loop is switched off. The differential conductance $dI/dV(V)$ is measured by sweeping the bias voltage between two values corresponding to two fixed energies.
The knowledge of the differential conductance gives access to the  local density of states (LDOS) of the sample.~\cite{Tersoff85} 

STM/STS techniques allow the study of individual SWNTs lying on a conductive substrate~\cite{buchs:245505,Ouyang:88.066804,Venema:1999p51,Lemay:2001p617} but also of SWNT devices with a source-drain and back-gate geometry.~\cite{Leroy:2007p3138} In SGM, the biased tip | playing the role of a local gate | is scanned at fixed height over a SWNT device between source and drain contacts.~\cite{Fan:2005p2662,Bockrath_Science01} The Fermi level of the tube is locally shifted due to charge accumulation near the tip and thus a map of the two-terminal transconductance $dI/dV_\text{g}$ of the nanotube device can be established. In all these cases, the tip couples weakly  to the SWNT. A strong coupling regime can be reached with a conductive atomic force microscope (AFM) tip brought in contact with the nanotube in order, \emph{e.g.}, to perform length-dependent two-terminal conductivity measurements.~\cite{depablo:2002}

In this paper, we first investigate theoretically the transport properties of strongly coupled SWNT devices | perfect source and drain contacts | with a third terminal, experimentally realized by a conductive tip in the weak coupling regime. In absence of translational invariance due to the presence of scattering centers, we develop a general formalism for studying electron scattering within the so-called extended \kp model. We account for scattering centers such as structural defects, tube ends or arbitrary random potentials. The approach leads to a general scattering matrix formulation of transport properties | current and differential conductance | of imperfect SWNTs. We apply our method to the case of a low temperature STM/STS experiment. In this set-up, individual finite-length metallic SWNTs dispersed on a gold substrate  show particle-in-a-box states between consecutive scattering centers. Our approach permits a fundamental understanding of the nature of the observed scattering patterns in reciprocal space.~\cite{buchs:245505}

The paper is organized in the following way: in Sec.~\ref{sec:two} we give an overview of the electronic properties of SWNTs in the \kp approximation as well as in the extended \kp  approximation  accounting for the presence of scattering centers; in Sec.~\ref{sec:three} we employ  a field theoretical approach and the extended \kp model to develop a method suitable to study the quantum transport properties of imperfect metallic SWNTs; in Sec.~\ref{sec:four} we describe the experimental setup,  the sample preparation, and the Fourier transform STS technique used for data analysis; in Sec.~\ref{sec:five} we compare the results of theory and experiment for the case of diverse SWNTs.

%%%%%%%%%%%%%%%%%%%%%%%%%%%%%%%%%%%
%%%%%%%%%%%%%%%%%%%%%%%%%%%%%%%%%%%
%%%%%%%%%%%%%%%%%%%%%%%%%%%%%%%%%%%
%%%%%%%%%%%%%%%%%%%%%%%%%%%%%%%%%%%
%
%  THE EXTENDED KP MODEL 
%
\section{The extended \emph{\lowercase{k$\cdot$p}} model}\label{sec:two}

A SWNT can be regarded | from the geometrical point of view | as a single-layer graphene strip rolled into a cylinder. Its structure is generally indexed by its chiral vector $\mathbf{C}_h$ defined by the circumferential vector that starts and ends on the same lattice site of the SWNT (see \textit{e.g.} Refs.~[\onlinecite{saito:1998.icp,ando:74.777}]). The circumferential vector can be expressed as a linear combination of the two basis vectors $\mathbf{a}_1=a_0(\sqrt{3}/2,1/2)$ and $\mathbf{a}_2=a_0(\sqrt{3}/2,-1/2)$ of the honeycomb lattice, where $a_0=  \sqrt{3}\,a_\text{CC}$ is the lattice constant ($a_\text{CC} \approx 0.142$~nm is the bond length between carbon atoms).  Thus, $\mathbf{C}_h=n_1 \mathbf{a}_1 + n_2\mathbf{a_2}$, and the geometry of a SWNT is completely defined by the pair of integers $(n_1,n_2)$ | the chiral indices. The diameter $d_\text{t}$ of a SWNT is given by:
%
%
%%%%%%%%%%%%%%%%
\begin{equation}
d_\text{t}=\frac{\left| \mathbf{C}_{h} \right|}{\pi}=\frac{a_0}{\pi} \sqrt{n_1^{2}+n_1 n_2+n_2^{2}}\,.
\end{equation}
%%%%%%%%%%%%%%%%
%
%
The unit cell of a SWNT is defined by the chiral vector $\mathbf{C}_{h}$ and the translational vector $\mathbf{T}$ perpendicular to $\mathbf{C}_{h}$. The translational vector $\mathbf{T}$ is the smallest lattice vector that leaves the SWNT lattice invariant when shifted by $\textbf{T}$ along the axis and its norm 
%
%
%%%%%%%%%%%%%%%%
\begin{equation}
    T=\left| \mathbf{T} \right|=\frac{\sqrt{3}a_0}{N_\text{R}}\sqrt{n_1^{2}+n_1 n_2+n_2^{2}}
\end{equation}
%%%%%%%%%%%%%%%%
%
%
defines the translational period. Here $N_\text{R}$ is the greatest common divisor of $(2n_2+n_1)$ and $(2n_1+n_2)$. The nanotube unit cell is thus formed by a cylindrical section of length $T$ and diameter $d_\text{t}$. It contains $N_\text{c}=4\left( n^{2}+nm+m^{2} \right)/N_\text{R}$  carbon atoms.

To understand the electronic properties of SWNTs, a simple way is to start with the band structure of graphene. This can be obtained from an effective tight-binding model with nearest-neighbor hopping with hopping integral $\gamma_0\approx 2.8\,$eV.~\cite{charlier:677}
The first Brillouin zone of the honeycomb lattice is characterized by the reciprocal lattice vectors $\mathbf{b}_1= (2\pi/a_0)(1/\sqrt{3},1)$ and $\mathbf{b}_2=(2\pi/a_0)(1/\sqrt{3},-1)$. Only two out of six vertices of the first Brillouin zone are non-equivalent | the $K$ and $K'$ points. A common choice is usually $\mathbf{K}=(2\pi/a_0)(1/\sqrt{3},1/2)$ and $\mathbf{K}'=(2\pi/a_0)(1/\sqrt{3},-1/2)$.
The conduction and valence bands formed by $\pi$-states touch at these $K$ and $K'$ points.  Due to periodic boundary conditions around the circumference of a SWNT, the allowed wave vectors perpendicular to the tube axis are quantized. In contrast, the wave vectors parallel to the tube axis remain continuous under the assumption of an infinitely long tube. The imposition of periodic boundary conditions in transverse direction leads to the following constraint on the electronic wavefunction:
%
%
%%%%%%%%%%%%%%%%
\begin{equation}\label{eq:pbc}
    \Psi_{\bm{k}}\left( \bm{r}+\mathbf{C}_h \right)=\ee^{\ii \bm{k} \cdot \mathbf{C}_h} \Psi_{\bm{k}} \left( \bm{r} \right) = \Psi_{\bm{k}} \left( \bm{r} \right)
\end{equation}
%%%%%%%%%%%%%%%%
%
%
where $\bm{r}$ is a vector on the tube surface and $\bm{k}$ lies in the first Brillouin zone of the nanotube.
Thus, the electronic states are restricted to $\bm{k}$ vectors that fulfill the condition:
%
%
%%%%%%%%%%%%%%%%%
\begin{equation}
    \bm{k}_\text{C} \cdot \mathbf{C}_h=2 \pi q
    \label{quantization}
\end{equation}
%%%%%%%%%%%%%%%%
%
%
with $q$ integer. Here $\bm{k}_\text{C}$ is the circumferential component of the electron momentum.  Plotting the endpoints of these allowed vectors for a given SWNT in the reciprocal space of graphene generates a series of parallel and equidistant lines. The distance between the lines is  $\Delta k_\text{C}=2/d_\text{t}$. The length, number and orientation of these lines depend on the chiral indices $(n_1,n_2)$ of the SWNT. If we restrict the wave vectors to the first Brillouin zone of the nanotube, its length is found to be $2\pi/T$ and the number of lines is equal to $N_\text{c}/2$.  The electronic band structure of a specific nanotube is given by the cuts of the graphene electronic energy bands along the corresponding allowed $\bm{k}$-lines: we find a pair of conduction (antibonding) and valence (bonding) bands for each $\bm{k}$-line (zone folding). Therefore, if the graphene $K$ and $K'$ points are crossed by an allowed $\bm{k}$-line, the SWNT is metallic, otherwise it is a semiconductor.

%
%
%%%%%%%%%%%%%%
\begin{figure}
	\begin{minipage}[c]{0.55\columnwidth}
		\includegraphics[width=0.95\textwidth]{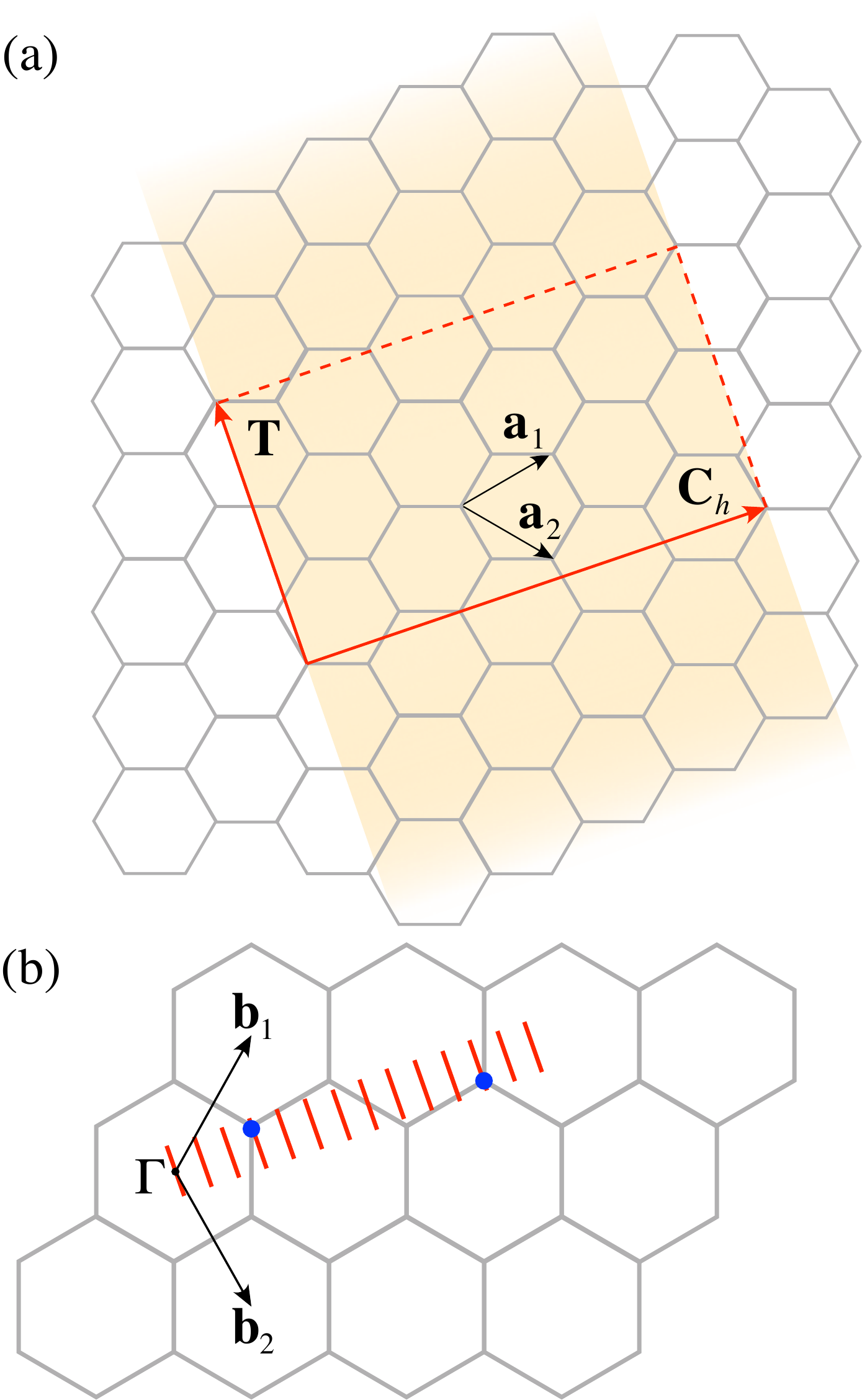}
	\end{minipage}
	\begin{minipage}[c]{0.40\columnwidth}
		\includegraphics[width=0.85\textwidth]{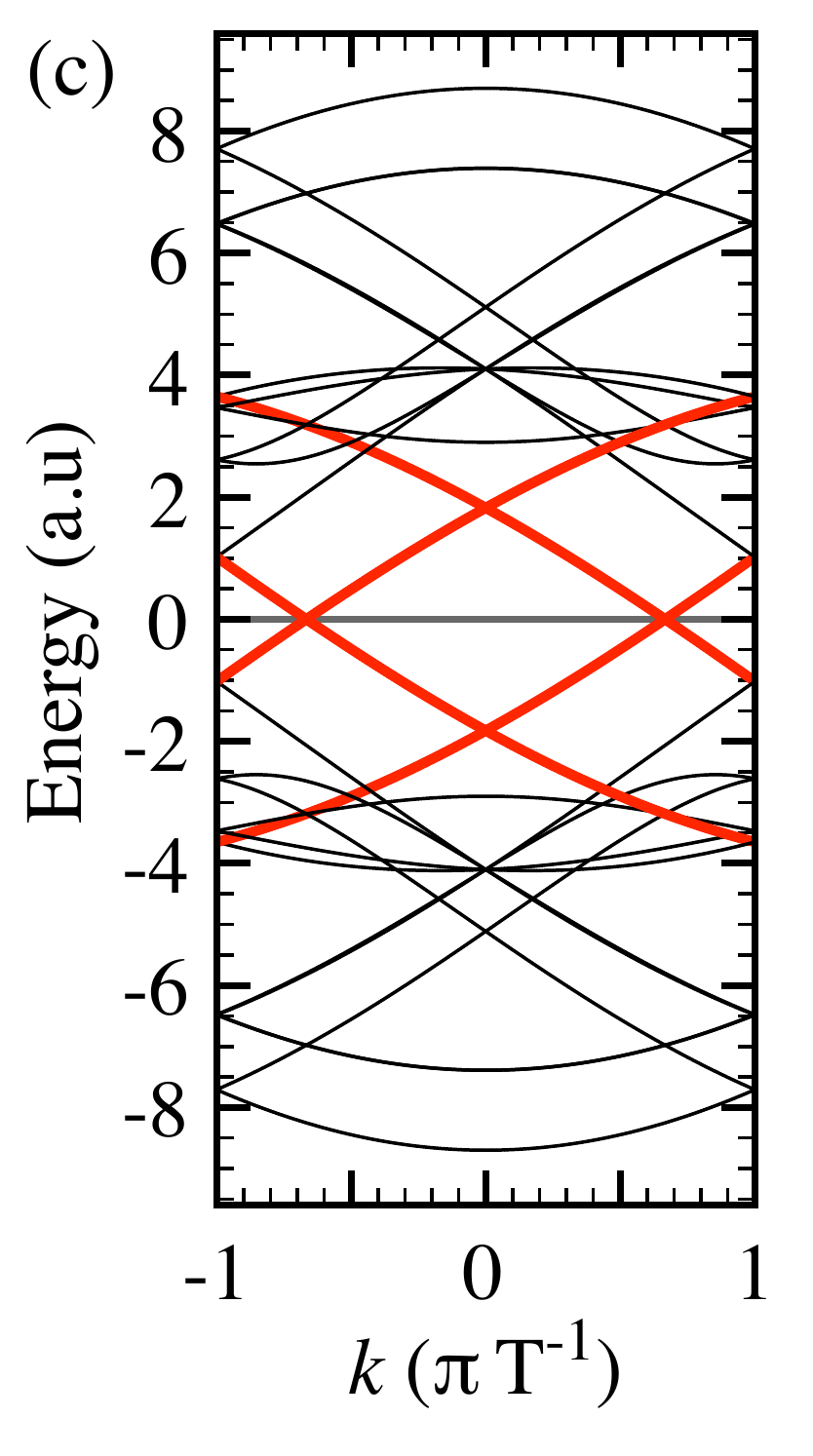}
	\end{minipage}
	\caption{\label{fig:one:a} (Color on-line). Panel~(a): the real space lattice of a carbon nanotube. The strip rolled into a cylinder is displayed as an orange  area. The vectors $\mathbf{a}_1$ and $\mathbf{a}_2$ are a basis vector set for the honeycomb lattice, $\mathbf{C}_h$ and $\mathbf{T}$ are the circumferential and the translational vector of the SWNT, respectively. Panel~(b): The reciprocal space of a carbon nanotube is represented by the red line segments. The figure correspond to a $(4,1)$ SWNT | a metallic tube. Here, $\mathbf{b}_1$ and $\mathbf{b}_2$ are the reciprocal lattice basis vectors for the honeycomb lattice. Panel~(c): Energy spectrum for the same SWNT. The red-thick lines show the low-energy bands investigated here.}
\end{figure}
%%%%%%%%%%%%%%
%
%
In Fig.~\ref{fig:one:a}(a) and (b) we show the case of a $(4,1)$ SWNT in the real and reciprocal space, respectively. Starting from the $\Gamma$-point we have a set of 14 parallel lines that are touching the vertices of the reciprocal honeycomb lattice in two points marked with dots | the $(4,1)$  is then a metallic SWNT.~\cite{note:two} In Fig.~\ref{fig:one:a}(c) we show the energy spectrum with energies measured relative to the CNP. The lines touching the dots representing $K$ and $K'$ of Fig.~\ref{fig:one:a}(b) are mapped in a set of bands (thick lines) crossing at the CNP. Around the CNP these bands have a quasi-linear behavior with respect to the momentum. Near the two $K$-points where the energy becomes very small, it is possible to perform an effective-mass or \kp approximation. Here, the wavefunction is factorized in a slowly-varying part | containing information about the lattice topology | times a plane wave. In this limit, the energy spectrum around each of the $K$ points  reads
%
%
%%%%%%%%%%%
\begin{equation}\label{eq:energy:kp}
\varepsilon_\pm(k) = \pm\hbar v_\text{F} k
\end{equation}
%%%%%%%%%%%
%
%
where $v_\text{F}$ is the effective Fermi velocity | the slope of the bands around zero energy. Here $k=|\bm{k}_\text{A}|$ is the longitudinal component of the  momentum relative to the $K$-points. The total longitudinal component of the electron momentum is $\alpha K\pm k$ ($\alpha=\pm$).
The energy spectrum~(\ref{eq:energy:kp}) presents a four-fold degenera\-cy: a two-fold one due to the presence of two independent $K$-points, and a two-fold one related to the electron spin. This  degeneracy leads to the fundamental result that the conductance at zero temperature is quantized and  equal to $G_0=4e^2/h$ for ideal ohmic contacts.

So far we have considered the properties of a perfect, infinitely long SWNT. However, this is the exception, in reality we usually are faced with tubes with imperfections, in particular defects in the atomic structure. These correspond in some cases to simple vacancies of one or more carbon atoms,~\cite{charlier:677} or lattice distorsions like Stone-Wales defects.~\cite{stone:128.501} Adatoms (C, H, N...) also modify the chemical structure of SWNTs.~\cite{Buchs_NJP_07, Buchs_Ar} In the presence of defects, the translational invariance of the SWNTs is broken. This implies that the electronic spectrum cannot simply be calculated by applying Bloch's theorem as done previously.

%
%
%%%%%%%%%%%%%%
\begin{figure}[t]
	\centering
	\includegraphics[width=0.6\columnwidth]{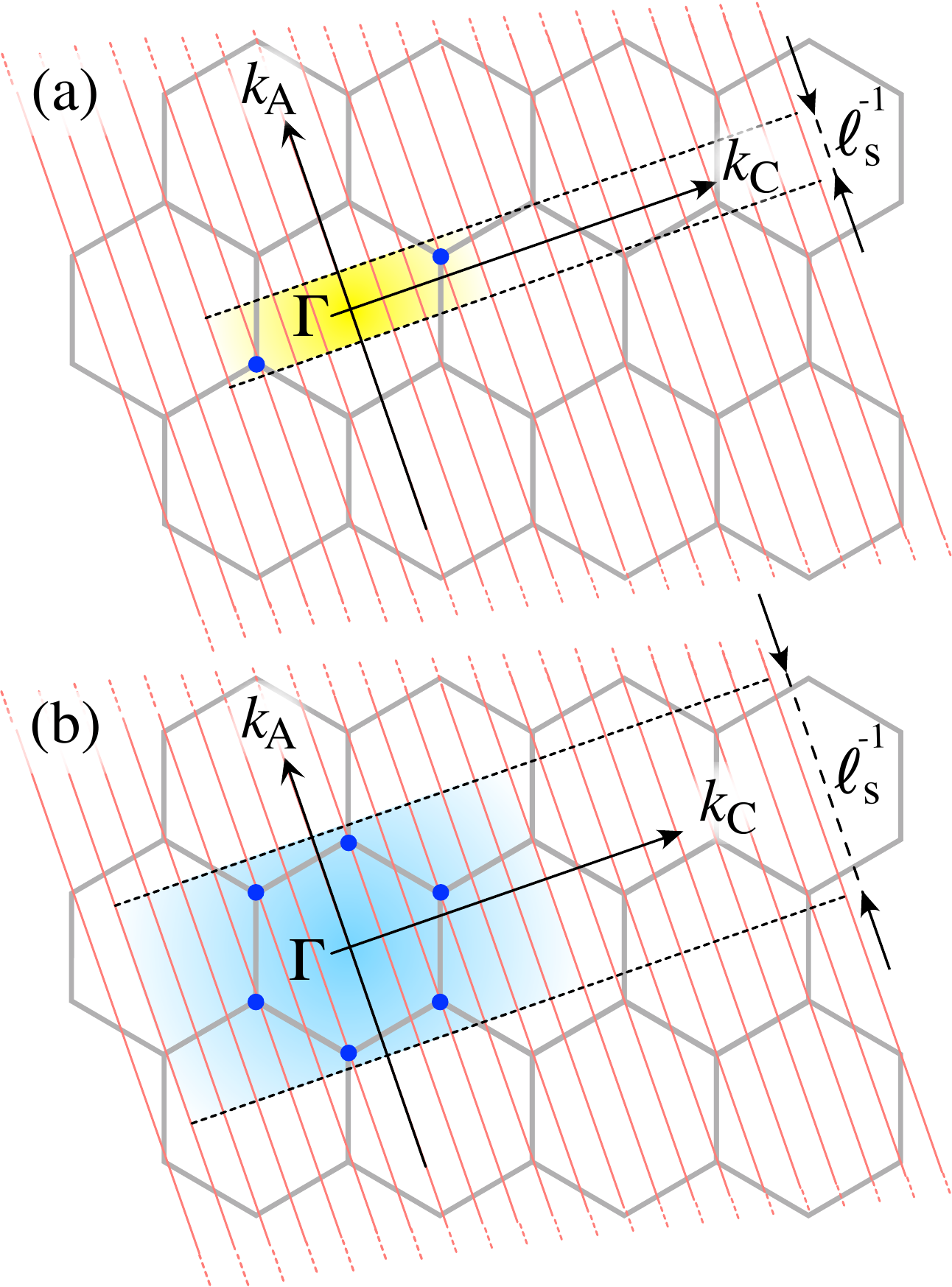}
	\caption{\label{figure_two} (Color on-line). Reciprocal space for a (4,1) SWNT in the case of an impurity breaking the translational invariance. Here $k_\text{A}$ and $k_\text{C}$ are the axial and circumferential axes, respectively. The effective size  $\ell_\text{s}$ of the scatterer in panel (a) is bigger than in panel (b).}
\end{figure}
%%%%%%%%%%%%%%
%
%

In the following we provide an approach tackling the problem of quantum transport in SWNTs in the presence of defects.
First we need to define the important length scales of the problem. 
A defect will distort the lattice structure of a SWNT in a region around the defect site characterized by a typical length, the effective size $\ell_\text{s}$ of the scatterer. This has to be compared with other characteristic lengths of SWNTs such as the diameter $d_\text{t}$, the translational period $T$, and the atomic lattice spacing $a_0$. In addition, defects and impurities  give rise to an effective scattering potential with a  potential strength $u_\text{s}$. This has to be compared with the effective hopping strength $\gamma_0$ between carbon atoms. 
In the following  we assume that the effective size $\ell_\text{s}$ of the scatterer is smaller than the SWNT diameter $d_\text{t}$, and that the impurity potential $u_\text{s}$ is smaller than the hopping strength $\gamma_0$.

Let us start with some considerations concerning the smallest of the characteristic length scale of a SWNT: the lattice constant $a_0$. In presence of long-range defects $\ell_\text{s} \gg a_0$  backscattering within the same valley or between different valleys is suppressed. This effect | known also as Klein tunneling | has been predicted in SWNTs for the first time in 1998 by Ando and Nakanishi~\cite{Ando:1998p1821} and has recently been observed in SWNT quantum dots.~\cite{steele:2009} On the other hand, if the effective size $\ell_\text{s}$ of  the defects is comparable to the lattice constant $a_0$, backscattering within the same valley | \emph{intra}-valley scattering | and between different valleys | \emph{inter}-valley scattering | become relevant. These scattering mechanisms become in general exponentially suppressed as the effective size $\ell_\text{s}$ of the defect grows.\cite{ando:74.777}

In order to understand how to go beyond the aforementioned  standard \kp model when dealing with defected SWNTs, we have to recall the role played by translational invariance in presence of umklapp processes. These processes are characterized by exchanged momenta exceeding in modulus the length of the reciprocal lattice vectors $\mathbf{G}$  | the electrons are scattered into a different Brillouin zone of the reciprocal lattice. For a  translationally invariant system, 
the equivalence of wave vectors differing by a reciprocal lattice vector allows to reduce these scattering events to the first Brillouin zone.~\cite{ando:74.777}

Defects change drastically this physical picture: the translational invariance is broken and wave vectors differing by reciprocal lattice vectors are no longer equivalent. We illustrate this situation in Fig.~\ref{figure_two}. Here we show the honeycomb reciprocal lattice and the parallel lines due to the periodic boundary conditions~(\ref{eq:pbc}).
In order to understand the consequences of the lack of translational invariance, we consider two distinct cases: $\ell_\text{s}\gg a_0$ in panel (a), and $a_0 \sim \ell_\text{s}<T$ in panel (b). The area of influence of defects in the reciprocal space is represented as a shadow area of approximative width $\ell_\text{s}^{-1}$. The precise extent of this area will depend on the details of the shape of the  Fourier transform of the defect potential. For a single defect, this is in general a decreasing function of the distance from the $\Gamma$ point.  In case (a), scattering events induced by the defect produce an exchanged momentum smaller than any reciprocal lattice vector. This is usually the regime in which backward scattering is suppressed |  therefore the scattering processes can all be studied within the original first Brillouin zone of the translationally invariant case. In this approximation we can adopt the conventional \kp picture. However, in case (b) this is no longer valid:
defects  induce scattering processes with exchanged momenta exceeding the size of the original first Brillouin zone. 
Hence, we have to account explicitly for all processes within a new extended scattering zone whose dimensions are given by the inverse effective size  $\ell_\text{s}^{-1}$ of the scatterer. In the \emph{extended} \kp approach we allow for new scattering channels that are associated with exchanged momenta larger than the difference between the original $K$ points. We refer to this consequences of broken translational symmetry as multi-component scattering. The exchanged momentum $\delta\bm{k}$ of each of these scattering processes will be characterized by an axial $\delta\bm{k}_\text{A}$ and a circumferential $\delta\bm{k}_\text{C}$ component, respectively (c.f. Fig.~\ref{fig_kp_ext}). This approach will be detailed in the following section.

%%%%%%%%%%%%%%%%%%%%%%%%%%%%%%%%%%%%%%%%%%
%%%%%%%%%%%%%%%%%%%%%%%%%%%%%%%%%%%%%%%%%%
%%%%%%%%%%%%%%%%%%%%%%%%%%%%%%%%%%%%%%%%%%
%
% Model Hamiltonian and scattering Processes
%
\section{Model Hamiltonian and scattering Processes}\label{sec:three}

We consider now a metallic SWNT device with ideal ohmic contacts to source (S) and drain (D) as well as a weak tunneling contact to a third  electrode. In our case this is an atomically sharp STM tip, henceforth denoted as tip (T). The SWNT has a finite length $L_0$ and for the $x$ coordinate along it we choose the origin in the
middle of the nanotube so that the interfaces to the S and D
electrodes are located at $x_\text{S}=-L_0/2$ and $x_\text{D}=+L_0/2$, respectively.
In the tube we account for the presence  of several defects characterized by an effective size $\ell_\text{s}$ comparable to the lattice constant $a_0$.
The tip is described as a semi-infinite non-interacting Fermi gas, and $y \le
0$ denotes the coordinate axis along the tip orthogonal to the
tube, the origin corresponding to the injection point on the tip.
The latter is located at position $x_\text{t}$ with respect to the middle
of the tube, and electron injection is modeled by a tunneling
Hamiltonian.
%
%
%%%%%%%%%%%%%%
\begin{figure}[t]
	\centering
	\includegraphics[width=0.8\columnwidth]{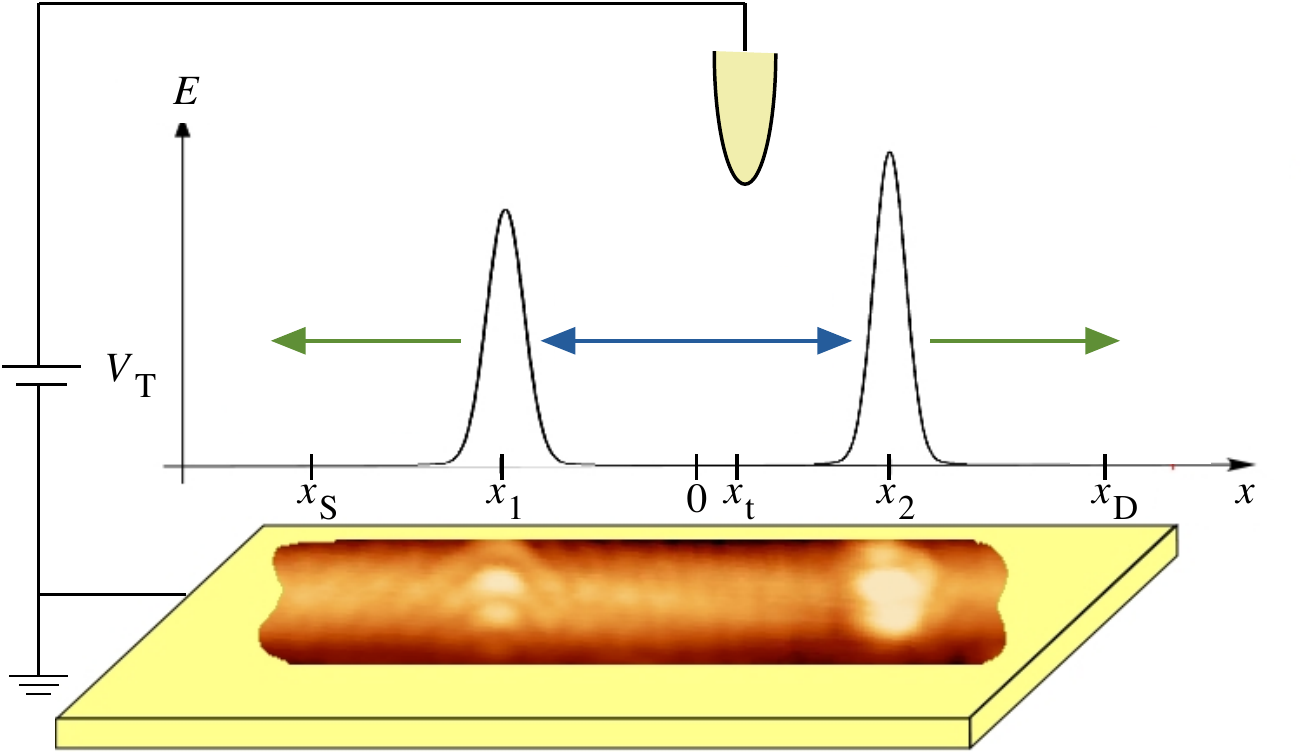}
	\caption{\label{fig:one} (Color on-line). Outline of the experimental set-up and of the Fabry-P\'erot resonator model.}
\end{figure}
%%%%%%%%%%%%%%
%
%
The total Hamiltonian of the system reads
%
%
%%%%%%%%%%%
\begin{equation}\label{eq:one}
\mathcal{H} = \mathcal{H}_\text{SWNT} + \mathcal{H}_\text{T} + \mathcal{H}_\text{tun}
\end{equation}
%%%%%%%%%%%
%
%
where the first term describes the defected SWNT and the coupling to the S and D leads. The second term accounts for the tip, whereas the last one includes SWNT-tip tunneling.

\begin{figure}[t]
	\centering
	\includegraphics[width=0.8\columnwidth]{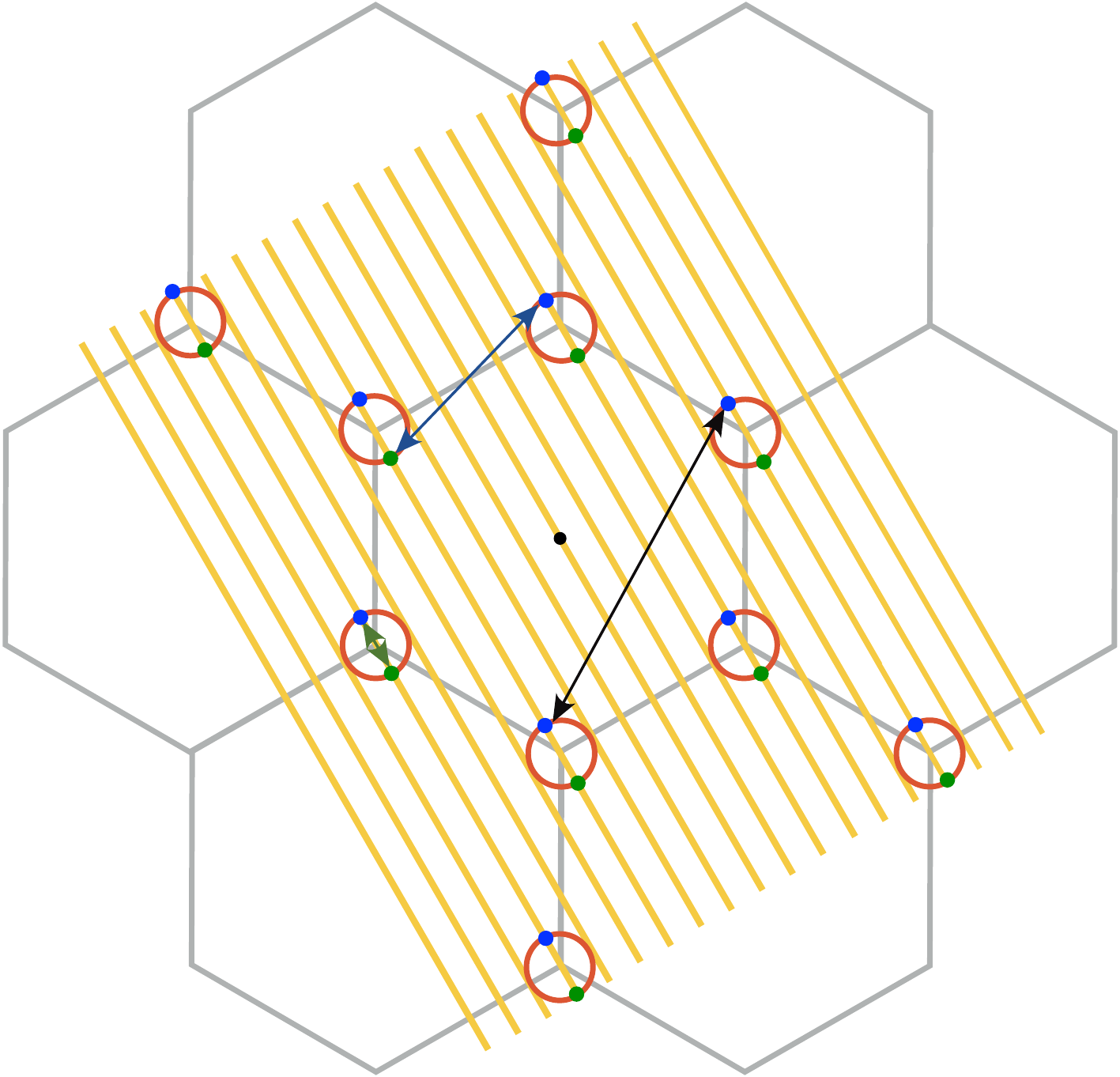}
	\caption{\label{fig_kp_ext} (Color on-line). Sketch of elastic scattering events by a local defect in a chiral (9,0) CNT within the extended \kp approximation. The parallel (yellow) lines are due to the periodic boundary condition along the tube circumference, their length is determined by the spatial extend of the defect. The (red) circle is the Fermi surface of the valleys at the energy of the scattered electrons. Available right [left] moving electron states are showed as green [blue] dots. The defect scatters electrons between all states indicated as dots. Some typical processes are showed explicitly. The green arrow shows an intra-valley backscattering event. The blue arrow depicts an inter-valley backscattering process. Finally, the black arrow indicates an extended inter-valley forward scattering event.}
\end{figure}

Concerning the SWNT, we focus on the low-energy excitations: the energy spectrum is considered as linear with respect to the electron momentum around the CNP (c.f.~Eq.~(\ref{eq:energy:kp})).  The electron operator $\Psi(x)$ can be decomposed around each of the two Dirac points $K,K'$ into right- and left-moving components $\Psi_{\qq,+}(x)$ and $\Psi_{\qq,-}(x)$, respectively.  The electron field reads
%
%
%%%%%%%%%%%
\begin{equation}\label{eq:three}
\Psi(x) = \sum_{\substack{\qq \in\{K,K'\} \\ \alpha=\pm}} \ee^{\ii (\qq +\alpha k) x}\Psi_{\qq\alpha}(x)
\end{equation}
%%%%%%%%%%%
%
%
where $k$ denotes the equilibrium Fermi momentum of the SWNT with respect to the charge neutrality point $\qq$. The Hamiltonian for the SWNT reads
%
%
%%%%%%%%%%%
\begin{equation}\label{eq:four}
\mathcal{H}_\text{SWNT} = \mathcal{H}_{\text{kin,SWNT}} + \mathcal{H}_\text{imp} + \mathcal{H}_{\mu_\text{SWNT}}\,.
\end{equation}
%%%%%%%%%%%
%
%
In the previous equation the first term
%
%
%%%%%%%%%%%
\begin{equation}\label{eq:five}
\mathcal{H}_\text{kin,SWNT} = -\ii \hbar \vf\sum_{\mathcal{Q},\alpha} \int_{-\infty}^{\infty} \!\!\! d x \,\alpha \no{\Psi^\dag_{\mathcal{Q}\alpha}(x) \partial_x \Psi_{\mathcal{Q}\alpha}(x)}
\end{equation}
%%%%%%%%%%%
%
%
describes the band linearized spectrum (\ref{eq:energy:kp}). The symbol $:\,\,:$ stands for normal ordering with respect to the equilibrium ground state. The second term in Eq.~\eqref{eq:four} models impurity scattering.
In the low-energy approximation, impurities with an effective size $\ell_\text{s}$ comparable to the lattice constant $a_0$ can be approximated by delta-like impurities. In order to fully account for the extended \kp model introduced in the previous section we have to account for multi-component scattering: Each of these impurities can either scatter electrons within the same valley | \emph{intra}-valley scattering | or can change the valley index of the backscattered electrons | \emph{inter}-valley scattering. 
In addition, we have to account for extended inter-valley scattering processes where the exchanged momenta are larger than the difference between the K points. 
In terms of field operators, the scattering Hamiltonian within the extended \kp model is of the form
%
%
%%%%%%%%%%%
\begin{align}\label{kpscattering}
\mathcal{H}_\text{imp} = & \hbar v_\text{F} \sum_n  \sum_{K,K'} \sum_G \sum_{\alpha,\alpha'} \lambda_{n}(\Delta K,\Delta\alpha,G) \\
&  \text{e}^{-\text{i} \left[\Delta K-G+ \Delta\alpha  k_\text{F}\right]x_n} \nonumber  \no{\Psi_{K\alpha}^\dag(x_n) \Psi^{}_{K'\alpha'}(x_n)}\,.
\end{align}
%%%%%%%%%%%
%
%
with $\Delta K= K-K'$ and $\Delta \alpha=\alpha-\alpha'$.
Here terms with $\alpha=\alpha'$ describe forward scattering processes and terms with $\alpha=-\alpha'$ describe backscattering. Moreover, for $K=K'$ we have intra-valley scattering while terms with $K\neq K'$  describe inter-valley scattering. Finally, $G$ is a reciprocal lattice vector and  $G\neq 0$ describes extended inter-valley scattering processes. For a particular case, the various scattering processes are sketched in Fig.~\ref{fig_kp_ext}.
In practice, the coupling constant $\lambda_n$  will depend on the Fourier transform of the scattering potential
 of the $n$--th defect. Hence, for a given defect  the coefficient $\lambda_n$ differs for each of the available scattering processes. 
Also the range of $G$-terms to be summed over depends on the spatial extend of the defects.
We will allow for several defects  at positions $x_1,\ldots,x_N$.

%
%
%%%%%%%%%%%%%%
%%%%%%%%%%%%%%%
%
%
The third term in Eq.~(\ref{eq:four}) accounts for the source and drain electrodes
%
%
%%%%%%%%%%%
\begin{equation}\label{eq:eight}
\mathcal{H}_{\mu_\text{SWNT}} = \int_{-\infty}^{\infty} dx \,\mu_\text{SWNT}(x) \rho(x)
\end{equation}
%%%%%%%%%%%
%
%
with
%
%
%%%%%%%%%%%
\begin{equation}\label{eq:nine}
\mu_\text{SWNT}(x) = \begin{cases}
e V_\s & \text{for}~~x<-x_\text{S} \\
e V_\D & \text{for}~~x>x_\text{D}
\end{cases}\,,
\end{equation}
%%%%%%%%%%%
%
%
where the term $\rho(x)=\no{\Psi^\dag(x) \Psi(x)}$ is the electron density fluctuation with respect to the equilibrium value.

The Hamiltonian of the tip, the second term in Eq.~(\ref{eq:one}),
reads
%
%
%%%%%%%%
\begin{equation}\label{eq:ten}
\mathcal{H}_{\T}= \mathcal{H}_\text{kin,T}+\mathcal{H}_{\mu_\T}\,.
\end{equation}
%%%%%%%%
%
%
Here
%
%
%%%%%%%%
\begin{equation}\label{eq:eleven}
\mathcal{H}_\text{kin,T}= -\ii \hbar v_\T \int_{-\infty}^{\infty}
\!\!\! dy \, \,\no{c^{\dagger}(y) \partial_y c^{}(y)}
\end{equation}
%%%%%%%%
%
%
describes the (linearized) band energy with respect to the
equilibrium Fermi points $\pm k_\T$ of the tip, and $v_\T$ denotes
the Fermi velocity. Notice that the integral runs also over the
positive $y$-axis, since right and left moving electron operators
along the physical tip axis $y<0$ have been unfolded into one
chiral (right-moving) operator $c^{}(y)$ defined on the whole
$y$-axis. The second term in Eq.~(\ref{eq:ten}) describes the bias
$V_\T$ applied to the tip that affects the incoming electrons
according to
%
%
%%%%%%%%%
\begin{equation}\label{eq:twelve}
\mathcal{H}_{\mu_\T}= e V_\T \int_{-\infty}^{0} \!\!\! dy \,
\no{c^{\dagger}(y) c^{}(y)} .
\end{equation}
%%%%%%%%%
%
%
Finally, the third term in Eq.~(\ref{eq:one}) accounts for the
SWNT-tip tunneling and reads
%
%
%%%%%%%%%
\begin{align}\label{eq:thirteen}
\mathcal{H}_\text{tun} = & \hbar \gamma \sqrt{v_\text{F} v_\T} ~\times \\
& \times \left[ \left( \sum_{\mathcal{Q},\alpha} \ee^{-\ii(\mathcal{Q}+\alpha k)x_\text{t}} \Psi_{\mathcal{Q}\alpha}^\dag(x_0) \right) c(0) + \text{h.c.} \right]
\nonumber
\end{align}
%%%%%%%%
%
%
where $\gamma$ is the tunneling amplitude for electrons, and $x=x_\text{t}$ [$y=0$] is the coordinate of the injection
point along the tube [tip].

In the following we will evaluate the differential conductance
in the tip terminal of the described set-up. Explicitly, due to the unfolding
procedure described above, the electron current flowing in the tip
at a point $y\le 0$ acquires the form
%
%
%%%%%%%%%%
\begin{equation}\label{eq:fourteen}
I_\text{T}(y,t) = e v_\T \left\langle \no{c^\dagger(y,t) c^{}(y,t)}
-\no{c^\dagger(-y,t) c^{}(-y,t)} \right\rangle\,.
\end{equation}
%%%%%%%%%%
%
%
Since the system Hamiltonian~(\ref{eq:one})  is quadratic in the fields
$\Psi_{\qq\alpha}(x)$ and $c(y)$ it  admits an exact solution. In fact, we can determine the
transport properties  within the Landauer-B\"uttiker formalism.~\cite{buettiker:1986,pugnetti:035121} In the
three-terminal set-up the energy-dependent scattering
matrix $\mathcal{S}(E)$ is a $5\times5$ matrix. It is a function of the tip position, the applied tip-SWNT bias, and the exchanged momenta $\delta \bm{k}$ in the scattering processes described by the extended \kp model. The axial component $\delta k_\text{A}$ of $\delta \bm{k}$ determines the momentum exchanged by the electrons  along their direction of motion. On the other side, the circumferential component $\delta k_\text{C}$ of $\delta \bm{k}$ affects | via the values of $\lambda_n$ |  the strength of the  scattering process.  In the following the scattering matrix  is expressed by the simplified notation
%
%
%%%%%%%%%%%
\begin{equation}\label{eq:sm:def}
\mathcal{S}_{\alpha,\beta} \equiv  
\mathcal{S}_{\alpha,\beta}(x_\text{t},E;\delta \bm{k})
\end{equation}
%%%%%%%%%%%
%
%
where the energy $E$ is measured with respect to the CNP.
The scattering matrix is obtained by solving the equations of motion for the system formed by the SWNT and the tip of the STM.
The current flowing through the tip from the SWNT may be written as
%
%
%%%%%%%%%%%
\begin{align}\label{eq:fifteen}
I_\text{T} =  \frac{2e}{h}  \int_{0}^{\infty} \!\!\! dE\ \Sigma(E)
\end{align}
%%%%%%%%%%%
%
%
with
%
%
%%%%%%%%%%%
\begin{align}\label{eq:sum:fun}
\Sigma(E) & = \left[  (|\mathcal{S}_{51}|^2+|\mathcal{S}_{53}|^2)(f_\text{T}(E)-f_\text{D}(E)) \right. \\
& \hspace{0.5cm} \left. +  (|\mathcal{S}_{52}|^2+|\mathcal{S}_{54}|^2)(f_\text{T}(E)-f_\text{S}(E)) \right]\,. \nonumber
\end{align}
%%%%%%%%%%%
%
%
Here $f_\text{S}(E)$, $f_\text{D}(E)$, and $f_\text{T}(E)$ are the Fermi distribution functions for the source, drain and tip contacts, respectively.

In the low temperature STM/STS experiment described in the next section, the differential conductance is deduced from the current flowing through the tip terminal.
In addition, in the experimental set-up  the metallic SWNT lays on a gold substrate, therefore its potential is uniform along the tube length. In our formalism this corresponds to the case of chemical potentials of source and drain electrodes fixed to the same value, so that $f_\text{S}(E)=f_\text{D}(E)$. Given these conditions,   the differential conductance $dI_\text{T}/dV_\text{T}$ can be readily obtained from  Eq.~(\ref{eq:fifteen}). At zero temperature we have
%
%
%%%%%%%%%%%
\begin{equation}\label{eq:sixteen}
\frac{dI_\text{T}}{dV_\text{T}} = \frac{2e^2}{h} \sum_{\ell=1}^{4} |\mathcal{S}_{5\ell}|^2\,.
\end{equation}
%%%%%%%%%%%
%
%
This is a function of the tip position and the voltage $V_\text{T}$ applied between the SWNT and the tip.

We want to add an important remark about the validity of the result (\ref{eq:sixteen}) for the differential conductance. The strong contact between the metallic SWNT and the gold substrate is ruling the phenomena observable experimentally | this leads to a short lifetime $\tau_\text{s}$ of  electrons injected from the tip into the SWNT.~\cite{rubio:1999:a,rubio:1999:b} Taking into account that experiments are performed at 4 K and that the applied voltages correspond to even larger energies, effects due to interactions with an energy scale smaller than $\hbar\tau_\text{s}^{-1}$ cannot be resolved in the local density of states.~\cite{eggert:2000}  In addition, the gold substrate is also partially screening the electronic Coulomb interaction in the SWNT and the tunneling junction resistance is dominating over the experimental set-up resistive circuit therefore excluding effects as the dynamical Coulomb blockade.~\cite{devoret:1990} Therefore, the only effect of  interaction that we take explicitly into account is a renormalization of the Fermi velocity~\cite{egger:1997} $v_\text{F}$ in Eq.~(\ref{eq:five}). 
Within this reasonable approximation the result (\ref{eq:sixteen}) is valid and will  suffice to describe the experimental observations.

%%%%%%%%%%%%%%%%%%%%%%%%
%%%%%%%%%%%%%%%%%%%%%%%%
%%%%%%%%%%%%%%%%%%%%%%%%
%
% THE EXPERIMENTAL SETUP
%
%
%
%%%%%%%%%%%%%%
\begin{figure*}
	\centering
	\includegraphics[width=0.7\textwidth]{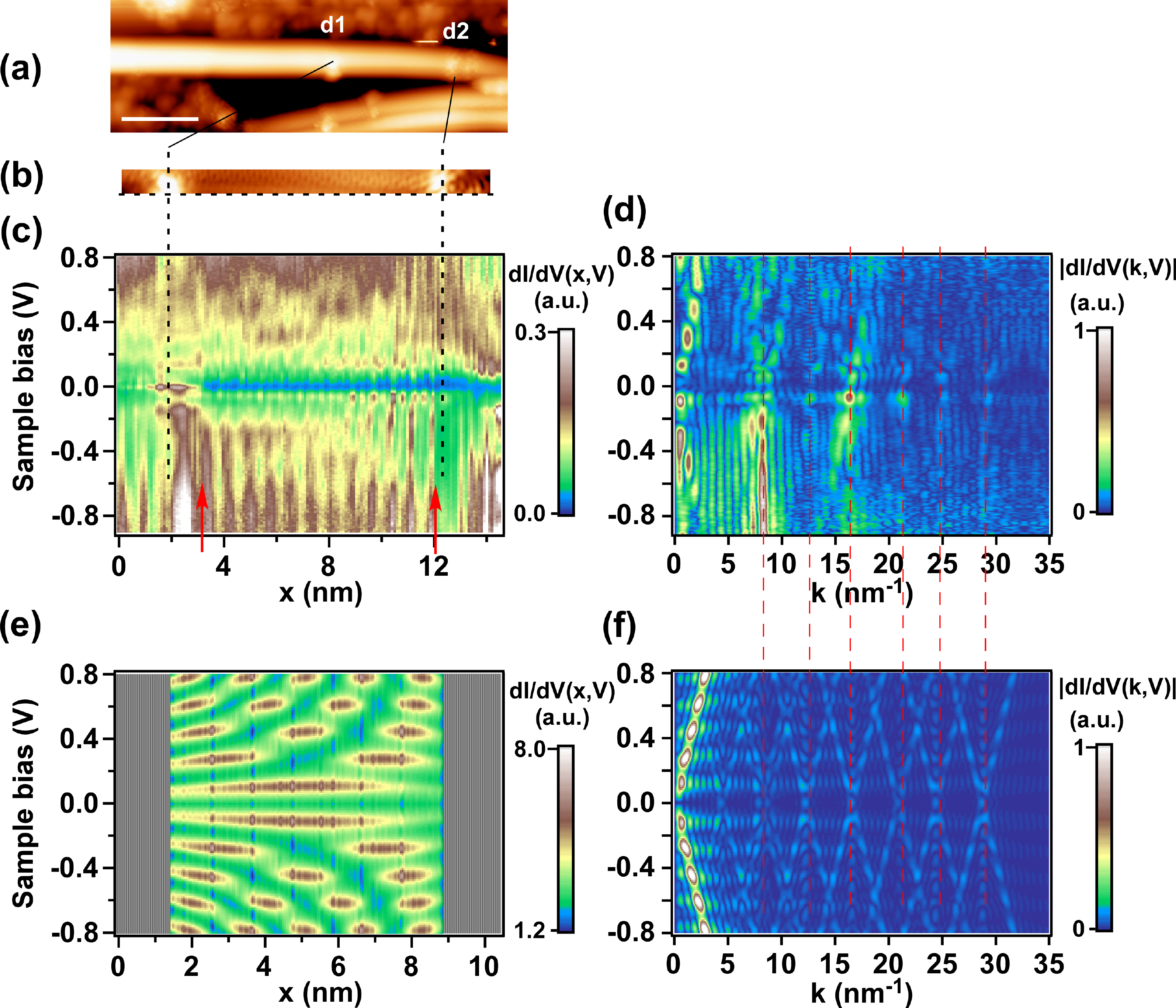}
	\caption{\label{exp_data_1} (Color on-line). (a) A portion of a metallic SWNT about $40$~nm long exposed to a low dose of 1.5 keV Ar$^{+}$ ions showing two defects labeled d1 and d2. (b) Detail image of the zone delimited by d1 and d2. (c) $dI/dV$-scan recorded along the horizontal dashed line in (b). (d) $\left| dI/dV(k,V) \right|^{2}$ map resulting from a line-by-line zero padding FFT performed between
the positions indicated by the two red arrows drawn in (c). (e) Calculated differential conductance for a (14,5) SWNT with inter-defect length $L = 10.4$~nm and asymmetric impurity strengths $\lambda_\text{R} = 0.35$ and $\lambda_\text{L} = 0.15$. (f) Corresponding $|dI/dV(k,V)|^2$ map.}
\end{figure*}
%%%%%%%%%%%%%%
%
%
\section{Experiments}\label{sec:four}

In the following section, we use our model to study the electronic interference patterns observed in the LDOS both in real and reciprocal spaces between artificially created scattering centers on metallic SWNTs deposited on a conductive substrate. In the next three sections we give a brief introduction to the experimental set-up and the data analysis technique.

\subsection{LT-STM/STS setup}

    The experiments have been carried out in a commercial LT-STM set-up (Omicron) operating in ultra high
    vacuum (UHV) with a base pressure below 10$^{-10}$~mbar. The STM topography images have been recorded in
    the constant current mode. In this set-up we refer to sample bias being the potential difference between the grounded sample and
    the tip. We used mechanically cut Pt/Ir tips for the topographic scans and the spectroscopy measurements,
    where the metallic nature of the tips has been regularly checked on the conductive substrate. Differential conductance $dI/dV$ spectra | proportional to the LDOS in first approximation~\cite{Tersoff85} | were measured using the lock-in technique. Under open loop conditions a low amplitude a.c.\ bias (5-15 mV at 600 Hz) is added to the sample bias and the resulting a.c. component of the tunneling current is detected. All the measurements have been carried out at a temperature of 5 K. All STM topography images presented in this work have been treated with
    the open source WSXM software.~\cite{WSXM}

\subsection{Sample preparation}

    Gold on glass substrates have been used to support the SWNTs for the measurements. In order to obtain
    clean Au (111) monoatomic terraces, the gold surface has been prepared by several \emph{in situ} sputtering and
    annealing cycles. The SWNTs provided from Rice University were grown by the high pressure CO disproportionation
    process (HiPco) and purified.~\cite{Smalley:8297}
    To maximize the number of individual nanotubes, the raw material was sonicated for 2-3 h in 1,2-dichloroethane
    and a droplet of the obtained SWNT suspension was then deposited \emph{ex situ} on the well prepared
    gold surface. Finally, residues from the solvent of the suspension were removed \emph{in situ} by means of a
    short sample annealing at about 390~$^{\circ}$C. More details on the sample preparation can be found in Ref.~[\onlinecite{Buchs_APL_07}].

    Strong electron scattering centers were created by two methods: bombardment with medium energy Ar$^{+}$ ions inducing vacancy-type defects,~\cite{GomezNavarro:2005p534,Tolvanen:2009} whereas voltage pulses in the STM created cuts of the SWNT at the tip position. In the former case, we used a standard Leybold\texttrademark \ ion gun designed to be integrated with  the preparation
    chamber of the STM setup. The ion gun and exposition parameters needed to achieve a defect density along the
    tube axis of about 0.1~nm$^{-1}$ have been obtained from irradiation tests performed on a freshly cleaved graphite (HOPG) substrate, as described in details in Ref.~[\onlinecite{Tolvanen:2009}].
    In order to create short SWNT segments, we have made use of the technique reported earlier by Lemay \emph{et~al.} in Ref.~[\onlinecite{Lemay_nat01}] where nanotubes were cut by applying a local voltage pulse of about $\left| V \right|= 5$~V at constant height (open loop).
It is known that the ultrasonication process used to disperse the nanotubes can also introduce defects. However, we checked the tubes before ion bombardment and found at most one defect every 200 nm. Since we aimed at a density of one defect per 10 nm, not more than one out of twenty defects was introduced by the ultrasonication process. The precise creation of a scattering defect is in fact not of much relevance for the electron scattering processes discussed here.

\subsection{Real and reciprocal space measurements}

    Electronic interference patterns induced between consecutive scattering centers are measured by
    recording a series of equidistant STS spectra along the SWNT.  In a first approximation, these give information about the LDOS along the tube axis.~\cite{Tersoff85} Typical $dI/dV (x, E(eV))$ data sets | called $dI/dV$-scans in the following | generally consist of 150 $dI/dV$ spectra recorded on topography line scans of 300 pts.

    Insights into the electron scattering leading to the interference patterns are obtained in the reciprocal space by means of the Fourier transform STS technique. Here, line-by-line zero-padding Fourier transforms of the $dI/dV$-scans are performed in the spatial region of interest | generally between two consecutive scattering centers | where the offset amplitude
    of each line has been subtracted. In order to avoid aliasing effects in the resulting $\left| dI/dV(k,V) \right|^{2}$
    data set, we always make sure to fulfill the sampling theorem. This requires that the highest spectral
    component of a signal must be smaller than $2 \pi f_\text{s}/2$, where $f_\text{s}$ is the spatial sampling frequency.~\cite{Shannon} This frequency is defined as $f_\text{s}=1/d_\text{s}$ where $d_\text{s}$ is the distance between two consecutive measurement points in the $dI/dV$-scan.
    For our purpose, we can show that for the usual spacing $d_\text{s}=0.1-0.13$ nm in a $dI/dV$-scan, the relevant spectral features with $k<25$~nm$^{-1}$ are guaranteed free of aliasing effects up to the maximum analyzed length of about 20~nm.~\cite{buchs:245505}

    Note that the spatial extent of a $dI/dV$-scan generally acquired within a time frame of about 30~min is
    always observed to be slightly larger (of the order of 5\%) than the spatial extent of the successive (or previous)
    topography lines acquired within about 0.5~s, due to a dependency of the piezoelectric voltage constant on the
    scan velocity. This effect has been systematically corrected by a compression of the $x$-scale of the
    $dI/dV$-scan to allow for a consistent comparison between the topographic and spatially resolved spectroscopic
    information on the same nanotube.

%%%%%%%%%%%%%%%%%%%%%%%%
%%%%%%%%%%%%%%%%%%%%%%%%
%%%%%%%%%%%%%%%%%%%%%%%%
%
% RESULTS
%
\section{Comparison between Theory and Experiment}\label{sec:five}

The first experimental example to which we applied our model is illustrated in Fig.~\ref{exp_data_1}.
Panel (a) shows the topography image of a $\sim30$~nm long portion of a metallic SWNT
that has been exposed to a low dose of $1.5$~keV Ar$^{+}$ ions.~\cite{Tolvanen_Ar,buchs:245505} A
detail image of the zone delimited by the pair of defects labeled d1 and d2 is displayed in (b).
Defects induced by medium energy Ar$^+$ ions appear typically as hillocks with an apparent height ranging from $0.5$~\AA \ to $4$~\AA \ and a lateral extension between $5$~\AA\ and $10$~\AA.
We have shown in an earlier work \cite{Tolvanen:2009}, by means of a detailed comparison of the spectroscopic signature of defect sites with predictions of \emph{ab}-initio simulations, that 1.5 keV Ar$^+$ ion bombardment of semiconducting SWNTs mainly gives rise to single and double vacancies, C-adatoms as well as combinations of these individual species. For metallic tubes the same defect structures are expected, but because of prohibitively high computational costs, the exact structure of defects in metallic SWNTs can hardly be determined with high accuracy. However, since the effective size $\ell_\text{s}$ of these defects is comparable to the lattice constant $a_0$, they can approximately be treated as delta-like impurities with a given scattering strength as presented in Sec. \ref{sec:three}. We will show later in this section that this simplified approach is accurate enough to interpret our experimental data to a large extent.
In Panel (c) of Fig.~\ref{exp_data_1} we shows the corresponding $dI/dV$-scan recorded along the horizontal dashed line in Panel (b). The
$\left| dI/dV(k,V) \right|^{2}$ map resulting from a line-by-line zero padding Fourier transform performed between
the positions indicated by the two red arrows drawn in (c) is depicted in (d). The low frequency Fourier
spots positioned between $k=0$ nm$^{-1}$ and $k=4$ nm$^{-1}$ correspond to intra-valley scattering processes.~\cite{buchs:245505} They are characterized by an average energy separation of $\Delta E =$160 meV. This is in
agreement with the value $\Delta E =$169 meV resulting from the one dimensional particle-in-a-box model
for the linear dispersion of a metallic nanotube ($E=\hbar v_\text{F} k$ around the inequivalent Fermi
points). The energy spacing can be obtained by
%
%
%%%%%%%%%%%%%%%%%
    \begin{equation}
        \Delta E = \hbar v_\text{F} \frac{\pi}{L} = \frac{h v_\text{F}}{2L} \simeq \frac{1.76}{L}\, \text{eV}~\text{nm}
        \label{deltaE}
    \end{equation}
%%%%%%%%%%%%%%%%%
%
%
with $L=10.4$~nm and the Fermi velocity $v_\text{F}= 8.5 \times 10^{5}$~m/s.~\cite{Lemay_nat01}

The more intense dispersive features allowing for
intervalley scattering are visible around $k \approx 8$ nm$^{-1}$ and at $k \approx 16$ nm$^{-1}$.
Further dispersive sets of spots are visible around $k=12$ nm$^{-1}$, $k=21$ nm$^{-1}$ , $k=25$ nm$^{-1}$ and $k=27$ nm$^{-1}$.
The left-right asymmetry of the interference pattern in the experimental $dI/dV$-scan (Fig. 5c) can be attributed to different scattering strengths of the defect sites d1 and d2.~\cite{buchs:245505} From atomically resolved topography 
images of the nanotube we can determine
a chiral angle of $\Theta \approx 16^\circ \pm 2^\circ$. Based on this value of the chiral angle and
the position of the Fourier spots in the $|dI/dV(k,V)|^2$ map, we find a good match with a
(14,5) metallic SWNT ($\Theta = 14.7^\circ$). 
Panel (e)  displays the calculated differential conductance
for a (14,5) SWNT with inter-defects length $L = 10.4$ nm and asymmetric impurity strengths $\lambda_\text{R} = 0.35$ and $\lambda_\text{L} = 0.15$. 
In general, each of the coefficients $\lambda_{n}(\Delta K,\Delta\alpha,G)$ in Eq.~\eqref{kpscattering} depends on the Fourier transformed scattering potential of the defect in question. Here we assume that this Fourier transform can adequately be described by a simple Gaussian.
The interference patterns show characteristic stripes with
increasing curvature when moving form the weaker left to the stronger right impurity.~\cite{buchs:245505}
All the observed Fourier components in (d) are reproduced in the calculated $|dI/
dV(k,V)|^2$ map in (f). 
The comparison between experiment and theory shows that regarding the positions and slopes of the scattering related branches close correspondence is found for the Fourier transformed $dI/dV$-scans. The intensity of the branches is, however, not well reproduced by the theory, which is the reason for a less pronounced agreement between the spatial $dI/dV$-maps. This is mainly due to the fact that in reality the scattering strength of a defect will depend on energy, momentum and pseudo-spin in a complex manner, which is not reproduced by the simplified delta-like scattering potential of our theoretical model. Nevertheless, the agreement between the Fourier transforms of the theoretical (f) and experimental (d) $dI/dV$-scans clearly proves that the model well reproduces the effect of multi-component scattering in metallic SWNTs. 
In the experiment, an asymmetry in the intensity of the positive and negative slope branches of the scattering process can be observed. 
In the experiment an asymmetry in the intensity of the positive and negative slope branches of the scattering process. The dispersion lines show a more intense signal for the positive slope branch than for the negative one. This asymmetry can be ascribed to a breaking of the particle-hole symmetry due to lattice reconstruction that mixes the two carbon sub-lattices of the SWNT structure at the defect site.~\cite{mayrhofer:2010} 

%
%
%%%%%%%%%%%%%%
\begin{figure}
	\centering
	\includegraphics[width=0.8\columnwidth]{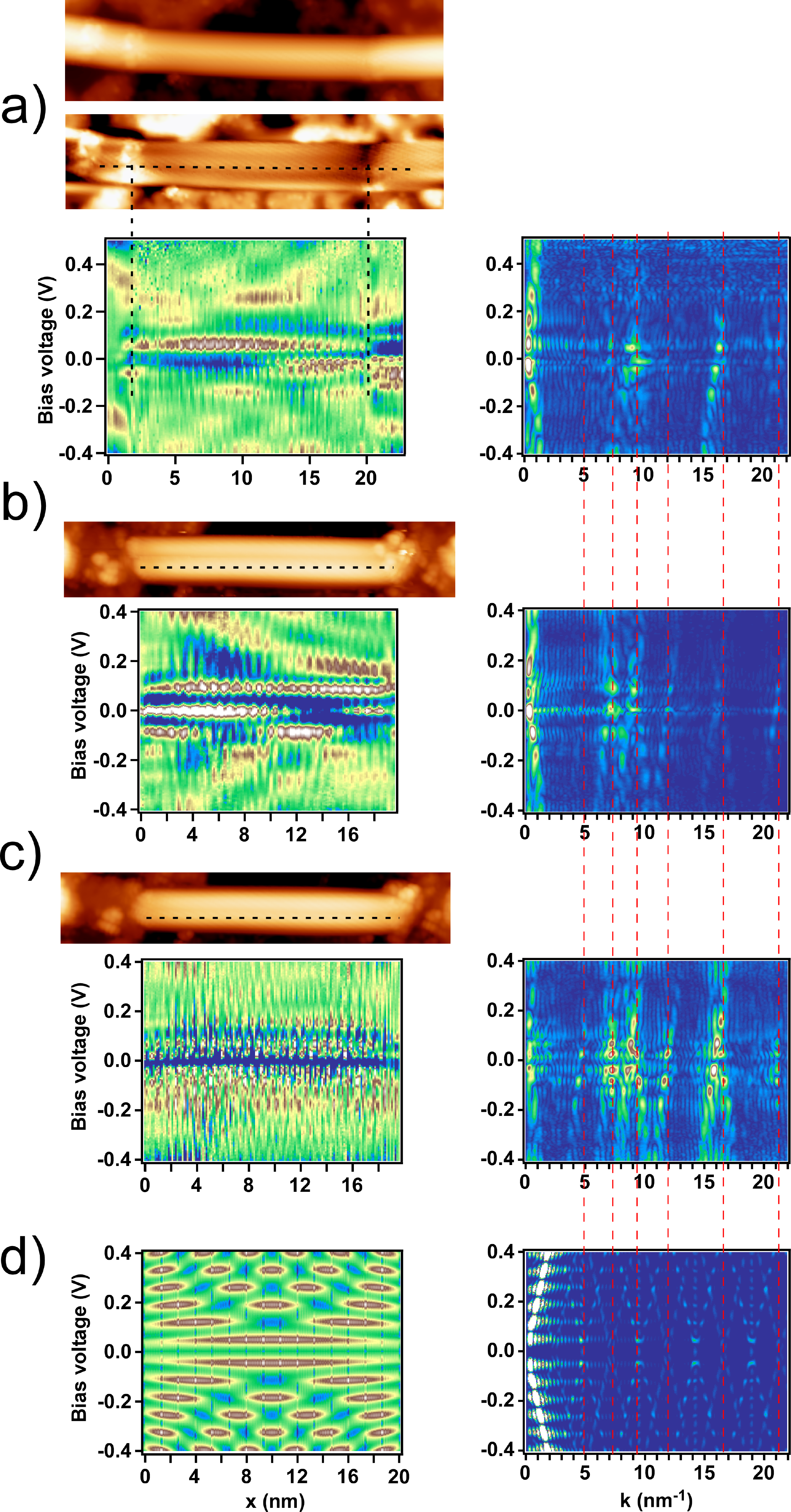}
	\caption{\label{exp_data_2} (Color on-line). STM topography image and corresponding $dI/dV$-scan of a metallic
SWNT exposed to 200 eV Ar$^{+}$ ions. (a) Situation with two defect sites. Upper panel shows a raw STM image of an about 30 nm long metallic SWNT with three defect sites, lower panel show the same data after plane correction and flattening.  (b) Situation
where two cuts have been performed at the defect positions by means of voltage
pulses at $ \approx \left |5\right |$~V. (c) A third voltage pulse has been performed close to the left
end. The color scale of the $|dI/dV(k,V)|^2$ map is the same for the three cases. $V_\text{s} = 0.5$~V, $I_\text{s} = 0.3$~nA, $T = 5.22$~K, $x_\text{res} = 0.15$~nm. (d) Calculated differential conductance with corresponding $|dI/dV(k,V)|^2$ map for a (10,4) SWNT with inter-defects length $L = 20$ nm and symmetric impurity strengths $\lambda_\text{R,L} = 0.3$.}
\end{figure}
%%%%%%%%%%%%%%
%
%
A second example is illustrated in Fig.~\ref{exp_data_2}. Panel (a) displays
a 23 nm long section of a metallic SWNT irradiated with 200~eV Ar$^{+}$ ions
showing two defect sites separated by $L \approx 17.5$~nm. The $dI/dV$-scan
shows  well-defined modes with an energy spacing of about
100~meV as expected from the separation of the two defects. In contrast to the example above, the interference
pattern  does not show curved stripes. This indicates that in this case the scattering strengths of the two defects  are similar. There is 
however a discernible symmetry breaking with respect to the $y$-axis at $x=L_{0}/2$,
which should be the position of a mirror-axis. This points to different scattering
phase shifts at the two defect sites. Note that our Fabry-P\'erot model does not
take the phase shift into account. When the tube is cut at both defect sites as
shown in panel (b), the mode pattern develops the typical asymmetric stripe pattern,
which we found to be characteristic for different scattering strengths (here with
$\lambda_\text{L} > \lambda_\text{R}$).~\cite{buchs:245505}
A further cutting pulse close to the left end of the tube results in the disappearance of the asymmetry
as shown in panel (c). It also leads to a weaker  resonant modes pattern,
indicating a reduction of the scattering strength at the left end of the tube. The left and right ends have now  similar scattering strengths.

This sequence of experiments indicates that even though one can expect tube ends to be
hard potential walls for the electrons, the scattering strengths at tube
ends can be different. This might result from different effects, \emph{e.g.},
due to different symmetry properties of the tube ends or due to a stronger tube--substrate interaction at the tube end. The latter can be induced by the cutting procedure,
leading to electron absorption instead of scattering. Furthermore, one can notice that
the $\left| dI/dV(k,V) \right|^{2}$ maps corresponding to the three cases (a), (b), and (c)
show intervalley scattering patterns with different intensities with some of them almost
suppressed. To allow a better analysis of these changes, we used our model to
calculate the differential conductance for a 20 nm long SWNT portion and the
corresponding $\left| dI/dV(k,V) \right|^{2}$ map for symmetric scattering strengths
at both ends (d). The best match was found for
a metallic (10,4) nanotube. Out of the nine projected Fermi cones included in the
extended \kp model [one right/left movers branch with a positive/negative slope
and a dispersiveless branch corresponding to forward scattering processes], about six cones
are visible in the $\left| dI/dV(k,V) \right|^{2}$ maps of the three cases presented in Fig.~\ref{exp_data_2}.
In each case, some of the cones are almost suppressed and some show a higher intensity. Also,
in this case we observe that some of the presented cones display only one dispersive slope. We can interpret these observation again in  terms of breaking of the particle-hole symmetry. In fact in the tube ends several hexagons are substituted by pentagons,~\cite{saito:1998.icp} this results in a direct bonding of carbon atoms on the same sub-lattice of the SWNT structure.~\cite{mayrhofer:2010}

%%%%%%%%%%%%%%%%%%%%%%%%
%%%%%%%%%%%%%%%%%%%%%%%%
%%%%%%%%%%%%%%%%%%%%%%%%
%
% DISCUSSION
%
\section{Conclusions}\label{sec:six}

We  have presented a general method for studying electron scattering  in metallic SWNTs in presence of defects. In particular, we focused on the presence of defects characterized by an effective size of the order of the SWNT lattice constant. 
In this limit, translational invariance is broken implying that the properties of SWNTs cannot be obtained from the Bloch theorem.
We have shown that  it is possible to introduce an extended \kp model in which the momentum exchanged in scattering processes can exceed the first Brillouin zone and cover an extended region depending on the effective size of defects. Within this scheme several scattering channels are available for electrons: intra- and inter-valley scattering in addition to extended inter-valley scattering. The latter becomes relevant because of the lack of translational  invariance. 

We have studied the consequences  of this kind of defects on the electronic properties of a  metallic SWNT contacted by  source and drain electrodes. Further, we have accounted for  a weak coupling to a tip. Based on a field theoretical approach describing low-energy electrons, we have obtained a general relation for the differential conductance between the SWNT and the tip, which provides access to the modifications of the local density of states induced by defects.
We have compared the results of our theoretical model for the local density of states with the results of experiments where | via the STM/STS technique | the same quantity has been measured  in defected or finite length metallic SWNTs. In the experimental set-up defects have been created via Ar$^+$ ion bombardment, while finite length tubes have been obtained by means of  the pulsed-voltage cutting technique.
The experimental results clearly show two important features of the theoretical approach. Firstly, the particle-in-a-box-like states observed experimentally between consecutive defects  are well described by a model with linear dispersion  for electrons close to the CNP. 
In accordance with the model, the measured level spacing  depends only on the distance between successive defects.
Secondly, the extended \kp approach fully succeeds in explaining the features observed in the Fourier transformed local density of states. This includes characteristics coming from  intra- and inter-valley scattering as well as these associated with extended inter-valley scattering.
Within this model the chiral angle determines uniquely the momenta exchanged in the scattering events causing the structure of the Fourier transformed local density of states. We have shown that the angle obtained from this procedure is in agreement with the one measured experimentally by atomically resolved topography images of SWNTs.

In spite of these assets, the  model for the impurities we have introduced in our theory fails in catching one essential feature observed experimentally in the Fourier transformed density of states. Specifically, we have routinely  observed an enhancement of some scattering channel with respect to others. These features can be explained by a more elaborate  atomistic description of defects accounting for a possible breaking of the particle-hole symmetry due to lattice reconstruction at the defect sites.~\cite{mayrhofer:2010} 

%%%%%%%%%%%%%%%%%%%%%%%%
%%%%%%%%%%%%%%%%%%%%%%%%
%%%%%%%%%%%%%%%%%%%%%%%%
%
% ACKNOWLEDGMENTS
%
\acknowledgments
We gratefully acknowledge helpful discussions with  A.~De Martino, M.~Grifoni, P. Gr\"oning, O.~Johnsen,  S.~G.~Lemay, L.~Mayrhofer, T.~Nakanishi, Y.~Nazarov, D.~Passerone, C.~Pignedoli, P.~Ruffieux and  Ch.~Sch\"onenberger. OG acknowledges financial support by the CabTuRes project funded by Nano-Tera.ch. The work of DB and HG is supported by the DFG grant BE~4564/1-1 and by the Excellence Initiative of the German Federal and State Governments.

%%%%%%%%%%%%%%%%%%%%%%%%
%%%%%%%%%%%%%%%%%%%%%%%%
%%%%%%%%%%%%%%%%%%%%%%%%


\begin{thebibliography}{45}
\expandafter\ifx\csname natexlab\endcsname\relax\def\natexlab#1{#1}\fi
\expandafter\ifx\csname bibnamefont\endcsname\relax
  \def\bibnamefont#1{#1}\fi
\expandafter\ifx\csname bibfnamefont\endcsname\relax
  \def\bibfnamefont#1{#1}\fi
\expandafter\ifx\csname citenamefont\endcsname\relax
  \def\citenamefont#1{#1}\fi
\expandafter\ifx\csname url\endcsname\relax
  \def\url#1{\texttt{#1}}\fi
\expandafter\ifx\csname urlprefix\endcsname\relax\def\urlprefix{URL }\fi
\providecommand{\bibinfo}[2]{#2}
\providecommand{\eprint}[2][]{\url{#2}}

\bibitem{Iijima:1991} S.Iijima, Nature \textbf{354}, 56 (1991).

\bibitem[{\citenamefont{Iijima and Ichihashi}(1993)}]{IIJIMA:1993.737}
\bibinfo{author}{\bibfnamefont{S.}~\bibnamefont{Iijima}} \bibnamefont{and}
  \bibinfo{author}{\bibfnamefont{T.}~\bibnamefont{Ichihashi}},
  \bibinfo{journal}{Nature} \textbf{\bibinfo{volume}{364}},
  \bibinfo{pages}{737} (\bibinfo{year}{1993}).

\bibitem[{\citenamefont{Bethune et~al.}(1993)\citenamefont{Bethune, Kiang,
  Devries, Gorman, Savoy, Vazquez, and Beyers}}]{BETHUNE:1993.605}
\bibinfo{author}{\bibfnamefont{D.}~\bibnamefont{Bethune}},
  \bibinfo{author}{\bibfnamefont{C.}~\bibnamefont{Kiang}},
  \bibinfo{author}{\bibfnamefont{M.}~\bibnamefont{Devries}},
  \bibinfo{author}{\bibfnamefont{G.}~\bibnamefont{Gorman}},
  \bibinfo{author}{\bibfnamefont{R.}~\bibnamefont{Savoy}},
  \bibinfo{author}{\bibfnamefont{J.}~\bibnamefont{Vazquez}}, \bibnamefont{and}
  \bibinfo{author}{\bibfnamefont{R.}~\bibnamefont{Beyers}},
  \bibinfo{journal}{Nature} \textbf{\bibinfo{volume}{363}},
  \bibinfo{pages}{605} (\bibinfo{year}{1993}).

\bibitem[{\citenamefont{Saito et~al.}(1998)\citenamefont{Saito, Dresselhaus,
  and Dresselhaus}}]{saito:1998.icp}
\bibinfo{author}{\bibfnamefont{R.}~\bibnamefont{Saito}},
  \bibinfo{author}{\bibfnamefont{G.}~\bibnamefont{Dresselhaus}},
  \bibnamefont{and}
  \bibinfo{author}{\bibfnamefont{M.}~\bibnamefont{Dresselhaus}},
  \emph{\bibinfo{title}{Physical Properties of Carbon Nanotubes}}
  (\bibinfo{publisher}{Imperial College Press, London}, \bibinfo{year}{1998}).

\bibitem[{\citenamefont{Toman\'ek and Enbody}(1999)}]{tomanek:1999.ka}
\bibinfo{author}{\bibfnamefont{D.}~\bibnamefont{Toman\'ek}} \bibnamefont{and}
  \bibinfo{author}{\bibfnamefont{R.}~\bibnamefont{Enbody}},
  \emph{\bibinfo{title}{Science and Application of Nanotubes}}
  (\bibinfo{publisher}{Kluwer Academic, Dordrecht}, \bibinfo{year}{1999}).

\bibitem[{\citenamefont{Ando}(2005)}]{ando:74.777}
\bibinfo{author}{\bibfnamefont{T.}~\bibnamefont{Ando}}, \bibinfo{journal}{J.
  Phys. Soc. Jpn.} \textbf{\bibinfo{volume}{74}}, \bibinfo{pages}{777}
  (\bibinfo{year}{2005}).

\bibitem[{\citenamefont{Charlier et~al.}(2007)\citenamefont{Charlier, Blase,
  and Roche}}]{charlier:677}
\bibinfo{author}{\bibfnamefont{J.-C.} \bibnamefont{Charlier}},
  \bibinfo{author}{\bibfnamefont{X.}~\bibnamefont{Blase}}, \bibnamefont{and}
  \bibinfo{author}{\bibfnamefont{S.}~\bibnamefont{Roche}},
  \bibinfo{journal}{Rev. Mod. Phys.} \textbf{\bibinfo{volume}{79}},
  \bibinfo{eid}{677} (\bibinfo{year}{2007}).

\bibitem[{\citenamefont{Steele et~al.}(2009{\natexlab{a}})\citenamefont{Steele,
  Hüttel, Witkamp, Poot, Meerwaldt, Kouwenhoven, and van~der
  Zant}}]{Steele:2009p1103}
\bibinfo{author}{\bibfnamefont{G.~A.} \bibnamefont{Steele}},
  \bibinfo{author}{\bibfnamefont{A.~K.} \bibnamefont{Hüttel}},
  \bibinfo{author}{\bibfnamefont{B.}~\bibnamefont{Witkamp}},
  \bibinfo{author}{\bibfnamefont{M.}~\bibnamefont{Poot}},
  \bibinfo{author}{\bibfnamefont{H.~B.} \bibnamefont{Meerwaldt}},
  \bibinfo{author}{\bibfnamefont{L.~P.} \bibnamefont{Kouwenhoven}},
  \bibnamefont{and} \bibinfo{author}{\bibfnamefont{H.~S.~J.}
  \bibnamefont{van~der Zant}}, \bibinfo{journal}{Science}
  \textbf{\bibinfo{volume}{325}}, \bibinfo{pages}{1103}
  (\bibinfo{year}{2009}{\natexlab{a}}).

\bibitem[{\citenamefont{Lassagne et~al.}(2009)\citenamefont{Lassagne,
  Tarakanov, Kinaret, Garcia-Sanchez, and Bachtold}}]{Lassagne:2009p1107}
\bibinfo{author}{\bibfnamefont{B.}~\bibnamefont{Lassagne}},
  \bibinfo{author}{\bibfnamefont{Y.}~\bibnamefont{Tarakanov}},
  \bibinfo{author}{\bibfnamefont{J.}~\bibnamefont{Kinaret}},
  \bibinfo{author}{\bibfnamefont{D.}~\bibnamefont{Garcia-Sanchez}},
  \bibnamefont{and} \bibinfo{author}{\bibfnamefont{A.}~\bibnamefont{Bachtold}},
  \bibinfo{journal}{Science} \textbf{\bibinfo{volume}{325}},
  \bibinfo{pages}{1107} (\bibinfo{year}{2009}).

\bibitem[{\citenamefont{Jarillo-Herrero
  et~al.}(2004)\citenamefont{Jarillo-Herrero, Sapmaz, Dekker, Kouwenhoven, and
  van~der Zant}}]{JarilloHerrero:2004p429}
\bibinfo{author}{\bibfnamefont{P.}~\bibnamefont{Jarillo-Herrero}},
  \bibinfo{author}{\bibfnamefont{S.}~\bibnamefont{Sapmaz}},
  \bibinfo{author}{\bibfnamefont{C.}~\bibnamefont{Dekker}},
  \bibinfo{author}{\bibfnamefont{L.~P.} \bibnamefont{Kouwenhoven}},
  \bibnamefont{and} \bibinfo{author}{\bibfnamefont{H.~S.~J.}
  \bibnamefont{van~der Zant}}, \bibinfo{journal}{Nature}
  \textbf{\bibinfo{volume}{389}}, \bibinfo{pages}{429} (\bibinfo{year}{2004}).

\bibitem{note:one} It has been recently shown that ultraclean metallic SWNT can show a Mott insulating phase. V.~V. Deshpande \emph{et al.}, Science \textbf{323}, 106 (2009).

\bibitem[{\citenamefont{Tomonaga}(1950)}]{TOMONAGA:1950.544}
\bibinfo{author}{\bibfnamefont{S.}~\bibnamefont{Tomonaga}},
  \bibinfo{journal}{Prog. Theor. Phys.} \textbf{\bibinfo{volume}{5}},
  \bibinfo{pages}{544} (\bibinfo{year}{1950}).

\bibitem[{\citenamefont{Luttinger}(1963)}]{LUTTINGER:1963.1154}
\bibinfo{author}{\bibfnamefont{J.}~\bibnamefont{Luttinger}},
  \bibinfo{journal}{J. Math. Phys.} \textbf{\bibinfo{volume}{4}},
  \bibinfo{pages}{1154} (\bibinfo{year}{1963}).

\bibitem[{\citenamefont{Egger and Gogolin}(1997)}]{Egger:1997.5082}
\bibinfo{author}{\bibfnamefont{R.}~\bibnamefont{Egger}} \bibnamefont{and}
  \bibinfo{author}{\bibfnamefont{A.}~\bibnamefont{Gogolin}},
  \bibinfo{journal}{Phys. Rev. Lett.} \textbf{\bibinfo{volume}{79}},
  \bibinfo{pages}{5082} (\bibinfo{year}{1997}).

\bibitem[{\citenamefont{Egger and Gogolin}(1998)}]{Egger:1998.281}
\bibinfo{author}{\bibfnamefont{R.}~\bibnamefont{Egger}} \bibnamefont{and}
  \bibinfo{author}{\bibfnamefont{A.}~\bibnamefont{Gogolin}},
  \bibinfo{journal}{Eur. Phys. J. B} \textbf{\bibinfo{volume}{3}},
  \bibinfo{pages}{281} (\bibinfo{year}{1998}).

\bibitem[{\citenamefont{Kane et~al.}(1997)\citenamefont{Kane, Balents, and
  Fisher}}]{PhysRevLett.79.5086}
\bibinfo{author}{\bibfnamefont{C.}~\bibnamefont{Kane}},
  \bibinfo{author}{\bibfnamefont{L.}~\bibnamefont{Balents}}, \bibnamefont{and}
  \bibinfo{author}{\bibfnamefont{M.~P.~A.} \bibnamefont{Fisher}},
  \bibinfo{journal}{Phys. Rev. Lett.} \textbf{\bibinfo{volume}{79}},
  \bibinfo{pages}{5086} (\bibinfo{year}{1997}).

\bibitem[{\citenamefont{Egger}(1999)}]{PhysRevLett.83.5547}
\bibinfo{author}{\bibfnamefont{R.}~\bibnamefont{Egger}},
  \bibinfo{journal}{Phys. Rev. Lett.} \textbf{\bibinfo{volume}{83}},
  \bibinfo{pages}{5547} (\bibinfo{year}{1999}).

\bibitem[{\citenamefont{Bockrath et~al.}(1999)\citenamefont{Bockrath, Cobden,
  Lu, Rinzler, Smalley, Balents, and McEuen}}]{Bockrath:1999.598}
\bibinfo{author}{\bibfnamefont{M.}~\bibnamefont{Bockrath}},
  \bibinfo{author}{\bibfnamefont{D.}~\bibnamefont{Cobden}},
  \bibinfo{author}{\bibfnamefont{J.}~\bibnamefont{Lu}},
  \bibinfo{author}{\bibfnamefont{A.}~\bibnamefont{Rinzler}},
  \bibinfo{author}{\bibfnamefont{R.}~\bibnamefont{Smalley}},
  \bibinfo{author}{\bibfnamefont{L.}~\bibnamefont{Balents}}, \bibnamefont{and}
  \bibinfo{author}{\bibfnamefont{P.}~\bibnamefont{McEuen}},
  \bibinfo{journal}{Nature} \textbf{\bibinfo{volume}{397}},
  \bibinfo{pages}{598} (\bibinfo{year}{1999}).

\bibitem[{\citenamefont{Yao et~al.}(1999)\citenamefont{Yao, Postma, and
  Dekker}}]{Yao:1999.273}
\bibinfo{author}{\bibfnamefont{Z.}~\bibnamefont{Yao}},
  \bibinfo{author}{\bibfnamefont{H.}~\bibnamefont{Postma}}, \bibnamefont{and}
  \bibinfo{author}{\bibfnamefont{C.}~\bibnamefont{Dekker}},
  \bibinfo{journal}{Nature} \textbf{\bibinfo{volume}{402}},
  \bibinfo{pages}{273} (\bibinfo{year}{1999}).

\bibitem[{\citenamefont{Gao et~al.}(2004)\citenamefont{Gao, Komnik, Egger,
  Glattli, and Bachtold}}]{PhysRevLett.92.216804}
\bibinfo{author}{\bibfnamefont{B.}~\bibnamefont{Gao}},
  \bibinfo{author}{\bibfnamefont{A.}~\bibnamefont{Komnik}},
  \bibinfo{author}{\bibfnamefont{R.}~\bibnamefont{Egger}},
  \bibinfo{author}{\bibfnamefont{D.~C.} \bibnamefont{Glattli}},
  \bibnamefont{and} \bibinfo{author}{\bibfnamefont{A.}~\bibnamefont{Bachtold}},
  \bibinfo{journal}{Phys. Rev. Lett.} \textbf{\bibinfo{volume}{92}},
  \bibinfo{pages}{216804} (\bibinfo{year}{2004}).

\bibitem[{\citenamefont{Postma et~al.}(2001)\citenamefont{Postma, Teepen, Yao,
  Grifoni, and Dekker}}]{Postma:293-76}
\bibinfo{author}{\bibfnamefont{H.~W.~C.} \bibnamefont{Postma}},
  \bibinfo{author}{\bibfnamefont{T.}~\bibnamefont{Teepen}},
  \bibinfo{author}{\bibfnamefont{Z.}~\bibnamefont{Yao}},
  \bibinfo{author}{\bibfnamefont{M.}~\bibnamefont{Grifoni}}, \bibnamefont{and}
  \bibinfo{author}{\bibfnamefont{C.}~\bibnamefont{Dekker}},
  \bibinfo{journal}{Science} \textbf{\bibinfo{volume}{293}},
  \bibinfo{pages}{76} (\bibinfo{year}{2001}).

\bibitem[{\citenamefont{Pugnetti et~al.}(2009)\citenamefont{Pugnetti, Dolcini,
  Bercioux, and Grabert}}]{pugnetti:035121}
\bibinfo{author}{\bibfnamefont{S.}~\bibnamefont{Pugnetti}},
  \bibinfo{author}{\bibfnamefont{F.}~\bibnamefont{Dolcini}},
  \bibinfo{author}{\bibfnamefont{D.}~\bibnamefont{Bercioux}}, \bibnamefont{and}
  \bibinfo{author}{\bibfnamefont{H.}~\bibnamefont{Grabert}},
  \bibinfo{journal}{Phys. Rev. B} \textbf{\bibinfo{volume}{79}},
  \bibinfo{eid}{035121} (\bibinfo{year}{2009}).

\bibitem[{\citenamefont{Liang et~al.}(2001)\citenamefont{Liang, Bockrath,
  Bozovic, Hafner, Tinkham, and Park}}]{Liang:2001p2177}
\bibinfo{author}{\bibfnamefont{W.}~\bibnamefont{Liang}},
  \bibinfo{author}{\bibfnamefont{M.}~\bibnamefont{Bockrath}},
  \bibinfo{author}{\bibfnamefont{D.}~\bibnamefont{Bozovic}},
  \bibinfo{author}{\bibfnamefont{J.~H.} \bibnamefont{Hafner}},
  \bibinfo{author}{\bibfnamefont{M.}~\bibnamefont{Tinkham}}, \bibnamefont{and}
  \bibinfo{author}{\bibfnamefont{H.}~\bibnamefont{Park}},
  \bibinfo{journal}{Nature} \textbf{\bibinfo{volume}{411}},
  \bibinfo{pages}{665} (\bibinfo{year}{2001}).

\bibitem[{\citenamefont{Wu et~al.}(2007)\citenamefont{Wu, Queipo, Nasibulin,
  Tsuneta, Wang, Kauppinen, and Hakonen}}]{wu:156803}
\bibinfo{author}{\bibfnamefont{F.}~\bibnamefont{Wu}},
  \bibinfo{author}{\bibfnamefont{P.}~\bibnamefont{Queipo}},
  \bibinfo{author}{\bibfnamefont{A.}~\bibnamefont{Nasibulin}},
  \bibinfo{author}{\bibfnamefont{T.}~\bibnamefont{Tsuneta}},
  \bibinfo{author}{\bibfnamefont{T.~H.} \bibnamefont{Wang}},
  \bibinfo{author}{\bibfnamefont{E.}~\bibnamefont{Kauppinen}},
  \bibnamefont{and} \bibinfo{author}{\bibfnamefont{P.~J.}
  \bibnamefont{Hakonen}}, \bibinfo{journal}{Phys. Rev. Lett.}
  \textbf{\bibinfo{volume}{99}}, \bibinfo{eid}{156803} (\bibinfo{year}{2007}).

\bibitem[{\citenamefont{Herrmann et~al.}(2007)\citenamefont{Herrmann, Delattre,
  Morfin, Berroir, Pla\c{c}ais, Glattli, and Kontos}}]{herrmann:156804}
\bibinfo{author}{\bibfnamefont{L.~G.} \bibnamefont{Herrmann}},
  \bibinfo{author}{\bibfnamefont{T.}~\bibnamefont{Delattre}},
  \bibinfo{author}{\bibfnamefont{P.}~\bibnamefont{Morfin}},
  \bibinfo{author}{\bibfnamefont{J.-M.} \bibnamefont{Berroir}},
  \bibinfo{author}{\bibfnamefont{B.}~\bibnamefont{Pla\c{c}ais}},
  \bibinfo{author}{\bibfnamefont{D.~C.} \bibnamefont{Glattli}},
  \bibnamefont{and} \bibinfo{author}{\bibfnamefont{T.}~\bibnamefont{Kontos}},
  \bibinfo{journal}{Phys. Rev. Lett.} \textbf{\bibinfo{volume}{99}},
  \bibinfo{eid}{156804} (\bibinfo{year}{2007}).

\bibitem[{\citenamefont{Tersoff and Hamann}(1985)}]{Tersoff85}
\bibinfo{author}{\bibfnamefont{J.}~\bibnamefont{Tersoff}} \bibnamefont{and}
  \bibinfo{author}{\bibfnamefont{D.~R.} \bibnamefont{Hamann}},
  \bibinfo{journal}{Phys. Rev. B} \textbf{\bibinfo{volume}{31}},
  \bibinfo{pages}{805} (\bibinfo{year}{1985}).

\bibitem[{\citenamefont{Buchs et~al.}(2009)\citenamefont{Buchs, Bercioux,
  Ruffieux, Gr\"{o}ning, Grabert, and Gr\"{o}ning}}]{buchs:245505}
\bibinfo{author}{\bibfnamefont{G.}~\bibnamefont{Buchs}},
  \bibinfo{author}{\bibfnamefont{D.}~\bibnamefont{Bercioux}},
  \bibinfo{author}{\bibfnamefont{P.}~\bibnamefont{Ruffieux}},
  \bibinfo{author}{\bibfnamefont{P.}~\bibnamefont{Gr\"{o}ning}},
  \bibinfo{author}{\bibfnamefont{H.}~\bibnamefont{Grabert}}, \bibnamefont{and}
  \bibinfo{author}{\bibfnamefont{O.}~\bibnamefont{Gr\"{o}ning}},
  \bibinfo{journal}{Phys. Rev. Lett.} \textbf{\bibinfo{volume}{102}},
  \bibinfo{eid}{245505} (\bibinfo{year}{2009}).

\bibitem[{\citenamefont{Ouyang et~al.}(2002)\citenamefont{Ouyang, Huang, and
  Lieber}}]{Ouyang:88.066804}
\bibinfo{author}{\bibfnamefont{M.}~\bibnamefont{Ouyang}},
  \bibinfo{author}{\bibfnamefont{J.-L.} \bibnamefont{Huang}}, \bibnamefont{and}
  \bibinfo{author}{\bibfnamefont{C.~M.} \bibnamefont{Lieber}},
  \bibinfo{journal}{Phys. Rev. Lett.} \textbf{\bibinfo{volume}{88}},
  \bibinfo{pages}{066804} (\bibinfo{year}{2002}).

\bibitem[{\citenamefont{Venema et~al.}(1999)\citenamefont{Venema, Wildoer,
  Janssen, Tans, Tuinstra, Kouwenhoven, and Dekker}}]{Venema:1999p51}
\bibinfo{author}{\bibfnamefont{L.}~\bibnamefont{Venema}},
  \bibinfo{author}{\bibfnamefont{J.}~\bibnamefont{Wildoer}},
  \bibinfo{author}{\bibfnamefont{J.}~\bibnamefont{Janssen}},
  \bibinfo{author}{\bibfnamefont{S.}~\bibnamefont{Tans}},
  \bibinfo{author}{\bibfnamefont{H.}~\bibnamefont{Tuinstra}},
  \bibinfo{author}{\bibfnamefont{L.}~\bibnamefont{Kouwenhoven}},
  \bibnamefont{and} \bibinfo{author}{\bibfnamefont{C.}~\bibnamefont{Dekker}},
  \bibinfo{journal}{Science} \textbf{\bibinfo{volume}{283}},
  \bibinfo{pages}{52} (\bibinfo{year}{1999}).

\bibitem[{\citenamefont{Lemay et~al.}(2001{\natexlab{a}})\citenamefont{Lemay,
  Janssen, van~den Hout, Mooij, Bronikowski, Willis, Smalley, Kouwenhoven, and
  Dekker}}]{Lemay:2001p617}
\bibinfo{author}{\bibfnamefont{S.}~\bibnamefont{Lemay}},
  \bibinfo{author}{\bibfnamefont{J.}~\bibnamefont{Janssen}},
  \bibinfo{author}{\bibfnamefont{M.}~\bibnamefont{van~den Hout}},
  \bibinfo{author}{\bibfnamefont{M.}~\bibnamefont{Mooij}},
  \bibinfo{author}{\bibfnamefont{M.}~\bibnamefont{Bronikowski}},
  \bibinfo{author}{\bibfnamefont{P.}~\bibnamefont{Willis}},
  \bibinfo{author}{\bibfnamefont{R.}~\bibnamefont{Smalley}},
  \bibinfo{author}{\bibfnamefont{L.}~\bibnamefont{Kouwenhoven}},
  \bibnamefont{and} \bibinfo{author}{\bibfnamefont{C.}~\bibnamefont{Dekker}},
  \bibinfo{journal}{Nature} \textbf{\bibinfo{volume}{412}},
  \bibinfo{pages}{617} (\bibinfo{year}{2001}{\natexlab{a}}).

\bibitem[{\citenamefont{Leroy et~al.}(2007)\citenamefont{Leroy, H., Pahilwani,
  , Dekker, and Lemay}}]{Leroy:2007p3138}
\bibinfo{author}{\bibfnamefont{B.~J.} \bibnamefont{Leroy}},
  \bibinfo{author}{\bibfnamefont{I.}~\bibnamefont{H.}},
  \bibinfo{author}{\bibfnamefont{V.~K.} \bibnamefont{Pahilwani}}, ,
  \bibinfo{author}{\bibfnamefont{C.}~\bibnamefont{Dekker}}, \bibnamefont{and}
  \bibinfo{author}{\bibfnamefont{S.~G.} \bibnamefont{Lemay}},
  \bibinfo{journal}{Nano Lett.} \textbf{\bibinfo{volume}{7}},
  \bibinfo{pages}{3138} (\bibinfo{year}{2007}).

\bibitem[{\citenamefont{Fan et~al.}(2005)\citenamefont{Fan, Goldsmith, and
  Collins}}]{Fan:2005p2662}
\bibinfo{author}{\bibfnamefont{Y.}~\bibnamefont{Fan}},
  \bibinfo{author}{\bibfnamefont{B.~R.} \bibnamefont{Goldsmith}},
  \bibnamefont{and} \bibinfo{author}{\bibfnamefont{P.~G.}
  \bibnamefont{Collins}}, \bibinfo{journal}{Nature Mater.}
  \textbf{\bibinfo{volume}{4}}, \bibinfo{pages}{906} (\bibinfo{year}{2005}).

\bibitem[{\citenamefont{Bockrath et~al.}(2001)\citenamefont{Bockrath, Liang,
  Bozovic, Hafner, Lieber, Tinkham, and Park}}]{Bockrath_Science01}
\bibinfo{author}{\bibfnamefont{M.}~\bibnamefont{Bockrath}},
  \bibinfo{author}{\bibfnamefont{W.}~\bibnamefont{Liang}},
  \bibinfo{author}{\bibfnamefont{D.}~\bibnamefont{Bozovic}},
  \bibinfo{author}{\bibfnamefont{J.~H.} \bibnamefont{Hafner}},
  \bibinfo{author}{\bibfnamefont{C.~M.} \bibnamefont{Lieber}},
  \bibinfo{author}{\bibfnamefont{M.}~\bibnamefont{Tinkham}}, \bibnamefont{and}
  \bibinfo{author}{\bibfnamefont{H.}~\bibnamefont{Park}},
  \bibinfo{journal}{Science} \textbf{\bibinfo{volume}{291}},
  \bibinfo{pages}{283} (\bibinfo{year}{2001}).

\bibitem{depablo:2002} J. de Pablo \emph{et al.}, Phys. Rev. Lett. \textbf{88}, 036804 (2002).

\bibitem{note:two} Effects of finite curvature open a gap in the energy spectrum of non-armchair SWNTs.

\bibitem[{\citenamefont{Stone and Wales}(1986)}]{stone:128.501}
\bibinfo{author}{\bibfnamefont{A.~J.} \bibnamefont{Stone}} \bibnamefont{and}
  \bibinfo{author}{\bibfnamefont{D.~J.} \bibnamefont{Wales}},
  \bibinfo{journal}{Chem. Phys. Lett.} \textbf{\bibinfo{volume}{128}},
  \bibinfo{pages}{501} (\bibinfo{year}{1986}).

\bibitem[{\citenamefont{Buchs et~al.}(2007{\natexlab{a}})\citenamefont{Buchs,
  Krasheninnikov, Ruffieux, Gr\"oning, Foster, Nieminen, and
  Gr\"oning}}]{Buchs_NJP_07}
\bibinfo{author}{\bibfnamefont{G.}~\bibnamefont{Buchs}},
  \bibinfo{author}{\bibfnamefont{A.~V.} \bibnamefont{Krasheninnikov}},
  \bibinfo{author}{\bibfnamefont{P.}~\bibnamefont{Ruffieux}},
  \bibinfo{author}{\bibfnamefont{P.}~\bibnamefont{Gröning}},
  \bibinfo{author}{\bibfnamefont{A.~S.} \bibnamefont{Foster}},
  \bibinfo{author}{\bibfnamefont{R.~M.} \bibnamefont{Nieminen}},
  \bibnamefont{and} \bibinfo{author}{\bibfnamefont{O.}~\bibnamefont{Gr\"oning}},
  \bibinfo{journal}{New J. Phys.} \textbf{\bibinfo{volume}{9}},
  \bibinfo{pages}{275} (\bibinfo{year}{2007}{\natexlab{a}}).

\bibitem[{\citenamefont{G\'omez-Navarro
  et~al.}(2005)\citenamefont{G\'omez-Navarro, Pablo, G\'omez-Herrero, Biel,
  Garcia-Vidal, Rubio, and Flores}}]{GomezNavarro:2005p534}
\bibinfo{author}{\bibfnamefont{C.~G.} \bibnamefont{G\'omez-Navarro}},
  \bibinfo{author}{\bibfnamefont{P.~J.~D.} \bibnamefont{Pablo}},
  \bibinfo{author}{\bibfnamefont{J.}~\bibnamefont{G\'omez-Herrero}},
  \bibinfo{author}{\bibfnamefont{B.}~\bibnamefont{Biel}},
  \bibinfo{author}{\bibfnamefont{F.~J.} \bibnamefont{Garcia-Vidal}},
  \bibinfo{author}{\bibfnamefont{A.}~\bibnamefont{Rubio}}, \bibnamefont{and}
  \bibinfo{author}{\bibfnamefont{F.}~\bibnamefont{Flores}},
  \bibinfo{journal}{Nature Mater.} \textbf{\bibinfo{volume}{4}},
  \bibinfo{pages}{534} (\bibinfo{year}{2005}).

\comment{\bibitem[{\citenamefont{Tolvanen
  et~al.}(2009{\natexlab{a}})\citenamefont{Tolvanen, Buchs, Krasheninnikov,
  Ruffieux, Gr\"oning, Foster, Nieminen, and Gr\"oning}}]{Buchs_Ar}
\bibinfo{author}{\bibfnamefont{A.}~\bibnamefont{Tolvanen}},
  \bibinfo{author}{\bibfnamefont{G.}~\bibnamefont{Buchs}},
  \bibinfo{author}{\bibfnamefont{A.~V.} \bibnamefont{Krasheninnikov}},
  \bibinfo{author}{\bibfnamefont{P.}~\bibnamefont{Ruffieux}},
  \bibinfo{author}{\bibfnamefont{P.}~\bibnamefont{Gr\"oning}},
  \bibinfo{author}{\bibfnamefont{A.~S.} \bibnamefont{Foster}},
  \bibinfo{author}{\bibfnamefont{R.~M.} \bibnamefont{Nieminen}},
  \bibnamefont{and}
  \bibinfo{author}{\bibfnamefont{O.}~\bibnamefont{Gr\"oning}},
  \bibinfo{journal}{Phys. Rev. B} \textbf{\bibinfo{volume}{79}},
  \bibinfo{pages}{125430} (\bibinfo{year}{2009}{\natexlab{a}}).}Ç

\bibitem[{\citenamefont{Ando and Nakanishi}(1998)}]{Ando:1998p1821}
\bibinfo{author}{\bibfnamefont{T.}~\bibnamefont{Ando}} \bibnamefont{and}
  \bibinfo{author}{\bibfnamefont{T.}~\bibnamefont{Nakanishi}},
  \bibinfo{journal}{J. Phys. Soc. Jpn.} \textbf{\bibinfo{volume}{67}},
  \bibinfo{pages}{1704} (\bibinfo{year}{1998}).

\bibitem[{\citenamefont{Steele et~al.}(2009{\natexlab{b}})\citenamefont{Steele,
  Gotz, and Kouwenhoven}}]{steele:2009}
\bibinfo{author}{\bibfnamefont{G.~A.} \bibnamefont{Steele}},
  \bibinfo{author}{\bibfnamefont{G.}~\bibnamefont{Gotz}}, \bibnamefont{and}
  \bibinfo{author}{\bibfnamefont{L.~P.} \bibnamefont{Kouwenhoven}},
  \bibinfo{journal}{Nature Nanotech.} \textbf{\bibinfo{volume}{4}},
  \bibinfo{pages}{363} (\bibinfo{year}{2009}{\natexlab{b}}).



\bibitem{rubio:1999:a} A.~Rubio \emph{et al.}, Phys. Rev. Lett. \textbf{82}, 3520 (1999).

\bibitem{rubio:1999:b} A.~Rubio, Appl. Phys. A \textbf{68}, 275 (1999).

\bibitem{eggert:2000} S. Eggert, Phys. Rev. Lett. \textbf{84}, 4413 (2000).


\bibitem{devoret:1990} M. H. Devoret \emph{et al.}, Phys. Rev. Lett. \textbf{64}, 1824 (1990).

\bibitem{egger:1997} R. Egger and H. Grabert, Phys. Rev. B \textbf{55}, 9929 (1997).

\bibitem[{\citenamefont{B\"uttiker}(1986)}]{buettiker:1986}
\bibinfo{author}{\bibfnamefont{M.}~\bibnamefont{B\"uttiker}},
  \bibinfo{journal}{Phys. Rev. Lett.} \textbf{\bibinfo{volume}{57}},
  \bibinfo{pages}{1761} (\bibinfo{year}{1986}).

\bibitem[{\citenamefont{Horcas et~al.}(2007)\citenamefont{Horcas, Fernandez,
  Gomez-Rodriguez, Colchero, J., and Baro}}]{WSXM}
\bibinfo{author}{\bibfnamefont{I.}~\bibnamefont{Horcas}},
  \bibinfo{author}{\bibfnamefont{R.}~\bibnamefont{Fernandez}},
  \bibinfo{author}{\bibfnamefont{J.~M.} \bibnamefont{Gomez-Rodriguez}},
  \bibinfo{author}{\bibfnamefont{J.}~\bibnamefont{Colchero}},
  \bibinfo{author}{\bibfnamefont{G.-H.} \bibnamefont{J.}}, \bibnamefont{and}
  \bibinfo{author}{\bibfnamefont{A.~M.} \bibnamefont{Baro}},
  \bibinfo{journal}{Rev. Sci. Instrum.} \textbf{\bibinfo{volume}{78}},
  \bibinfo{pages}{013705} (\bibinfo{year}{2007}).

\bibitem[{\citenamefont{Chiang et~al.}(2001)\citenamefont{Chiang, Brinson,
  Huang, Willis, Bronikowski, Margrave, Smalley, and Hauge}}]{Smalley:8297}
\bibinfo{author}{\bibfnamefont{I.~W.} \bibnamefont{Chiang}},
  \bibinfo{author}{\bibfnamefont{B.~E.} \bibnamefont{Brinson}},
  \bibinfo{author}{\bibfnamefont{A.~Y.} \bibnamefont{Huang}},
  \bibinfo{author}{\bibfnamefont{P.~A.} \bibnamefont{Willis}},
  \bibinfo{author}{\bibfnamefont{M.~J.} \bibnamefont{Bronikowski}},
  \bibinfo{author}{\bibfnamefont{J.~L.} \bibnamefont{Margrave}},
  \bibinfo{author}{\bibfnamefont{R.~E.} \bibnamefont{Smalley}},
  \bibnamefont{and} \bibinfo{author}{\bibfnamefont{R.~H.} \bibnamefont{Hauge}},
  \bibinfo{journal}{J. Phys. Chem. B} \textbf{\bibinfo{volume}{105}},
  \bibinfo{pages}{8297} (\bibinfo{year}{2001}).

\bibitem[{\citenamefont{Buchs et~al.}(2007{\natexlab{b}})\citenamefont{Buchs,
  Ruffieux, Gr\"oning, and Gr\"oning}}]{Buchs_APL_07}
\bibinfo{author}{\bibfnamefont{G.}~\bibnamefont{Buchs}},
  \bibinfo{author}{\bibfnamefont{P.}~\bibnamefont{Ruffieux}},
  \bibinfo{author}{\bibfnamefont{P.}~\bibnamefont{Gr\"oning}},
  \bibnamefont{and}
  \bibinfo{author}{\bibfnamefont{O.}~\bibnamefont{Gr\"oning}},
  \bibinfo{journal}{Appl. Phys. Lett.} \textbf{\bibinfo{volume}{90}},
  \bibinfo{pages}{013104} (\bibinfo{year}{2007}{\natexlab{b}}).

\bibitem[{\citenamefont{Lemay et~al.}(2001{\natexlab{b}})\citenamefont{Lemay,
  Janssen, van~den Hout, Mooij, Bronikowski, Willis, Smalley, Kouwenhoven, and
  Dekker}}]{Lemay_nat01}
\bibinfo{author}{\bibfnamefont{S.~G.} \bibnamefont{Lemay}},
  \bibinfo{author}{\bibfnamefont{J.~W.} \bibnamefont{Janssen}},
  \bibinfo{author}{\bibfnamefont{M.}~\bibnamefont{van~den Hout}},
  \bibinfo{author}{\bibfnamefont{M.}~\bibnamefont{Mooij}},
  \bibinfo{author}{\bibfnamefont{M.~J.} \bibnamefont{Bronikowski}},
  \bibinfo{author}{\bibfnamefont{P.~A.} \bibnamefont{Willis}},
  \bibinfo{author}{\bibfnamefont{R.~E.} \bibnamefont{Smalley}},
  \bibinfo{author}{\bibfnamefont{L.~P.} \bibnamefont{Kouwenhoven}},
  \bibnamefont{and} \bibinfo{author}{\bibfnamefont{C.}~\bibnamefont{Dekker}},
  \bibinfo{journal}{Nature} \textbf{\bibinfo{volume}{412}},
  \bibinfo{pages}{617} (\bibinfo{year}{2001}{\natexlab{b}}).

\bibitem[{\citenamefont{Shannon}(1949)}]{Shannon}
\bibinfo{author}{\bibfnamefont{C.~E.} \bibnamefont{Shannon}},
  \bibinfo{journal}{Proc. Institute of Radio Engineers}
  \textbf{\bibinfo{volume}{37}}, \bibinfo{pages}{10} (\bibinfo{year}{1949}).

\bibitem{Tolvanen:2009} A.~Tolvanen,
 G.~Buchs, P.~Ruffieux, P.~Gr\"oning, O.~Gr\"oning, and A.~V. Krasheninnikov, Phys. Rev. B
  \textbf{79},  125430 (2009).

\bibitem[{\citenamefont{Mayrhofer and Bercioux}(2010)}]{mayrhofer:2010}
\bibinfo{author}{\bibfnamefont{L.}~\bibnamefont{Mayrhofer}} \bibnamefont{and}
  \bibinfo{author}{\bibfnamefont{D.}~\bibnamefont{Bercioux}},
  \bibinfo{journal}{arXiv:1009.4839, \emph{unpublished}}  (\bibinfo{year}{2010}).

\end{thebibliography}
\end{document}